\pdfoutput=1
\documentclass{emulateapj}

\shorttitle{Multiwavelength Study of Sh2-297}
\shortauthors{Mallick et al.}

\usepackage{subfigure}
\usepackage{amsmath}
\usepackage{natbib}

\begin{document}

\title{Star formation activity in the Galactic H~{\sc ii} region Sh2-297}

\author{K. K. Mallick, D. K. Ojha}
\affil{Department of Astronomy and Astrophysics, Tata Institute of Fundamental Research, Homi Bhabha Road, 
       Colaba, Mumbai (Bombay) 400 005, India}
\email{kshitiz@tifr.res.in}

\author{M. R. Samal}
\affil{Laboratoire d'Astrophysique de Marseille (UMR 6110 CNRS \& Universit\'e de Provence), \\
       38 rue F. Joliot-Curie, 13388 Marseille Cedex 13, France}

\author{A. K. Pandey}
\affil{Aryabhatta Research Institute of Observational Sciences, Manora Peak, \\ 
       Nainital 263 129, India}

\author{B. C. Bhatt}
\affil{Indian Institute of Astrophysics, Koramangala, Bangalore 560 034, India}

\author{S. K. Ghosh}
\affil{National Centre for Radio Astrophysics, Tata Institute of Fundamental Research, \\
       Pune 411 007, India}

\author{L. K. Dewangan}
\affil{Department of Astronomy and Astrophysics, Tata Institute of Fundamental Research, Homi Bhabha Road, 
       Colaba, Mumbai (Bombay) 400 005, India}

\and

\author{M. Tamura}
\affil{National Astronomical Observatory of Japan, Mitaka, Tokyo 181-8588, Japan}

\begin{abstract}
We present a multiwavelength study of the Galactic H~{\sc ii} region
\object{Sh2-297}, located in Canis Major OB1 complex. Optical spectroscopic 
observations are used to constrain the spectral type of ionizing star 
\object{HD 53623} as B0V. The classical nature of this H~{\sc ii} region is  
affirmed by the low values of electron density and emission measure, 
which are calculated to be 756 cm$^{-3}$ and 9.15$\times$10$^{5}$ cm$^{-6}$ pc
using the radio continuum observations at 610 and 1280 MHz,
and VLA archival data at 1420 MHz.
To understand local star formation, we identified the young stellar object (YSO)
candidates in a region of area $\sim$ 7.5$^{'} \times$7.5$^{'}$ centered on \object{Sh2-297} 
using grism slitless spectroscopy (to identify the H$\alpha$ emission line stars), and
near infrared (NIR) observations. NIR YSO candidates are further classified 
into various evolutionary stages using color-color (CC) and color-magnitude (CM) diagrams, 
giving 50 red sources ($H-K>0.6$) and 26 Class II-like sources. 
The mass and age range of the YSOs are estimated to be $\sim$ 0.1$-$2 M$_{\odot}$ and  
0.5$-$2 Myr using optical (\textit{V/V-I}) and NIR (\textit{J/J-H}) CM diagrams. 
The mean age of the YSOs is found to be $\sim$ 1 Myr, which is of the order of dynamical 
age of 1.07 Myr of the H~{\sc ii} region. 
Using the estimated range of visual extinction (1.1$-$25 mag) from literature and 
NIR data for the region, 
spectral energy distribution (SED) models have been 
implemented for selected YSOs which 
show masses and ages to be consistent with estimated values. 
The spatial distribution of YSOs shows an evolutionary sequence,  
suggesting triggered star formation in the region. The star 
formation seems to have propagated from the ionizing star towards the  
cold dark cloud \object{LDN1657A} located west of \object{Sh2-297}.   
\end{abstract}


\section{Introduction}

The study of massive stars ($\gtrsim$ 8 M$_{\odot}$) 
is one of the most prolific areas
of research in present day astronomy, a fact which can be attributed
to the not-so-well understood processes and a range of theories 
\citep[see][and references therein]{zin07}. 
Massive star formation, though a rare event in itself, can have a
significant impact on its natal environment through processes such as strong stellar wind 
and supernovae explosions, which transfer large energy and momentum 
to the surrounding environment. 
Features such as bright-rimmed clouds and H~{\sc ii} regions (owing to 
their large output of UV photons) are hallmarks of massive stars in a region. 

\begin{figure}
\includegraphics[scale=0.4]{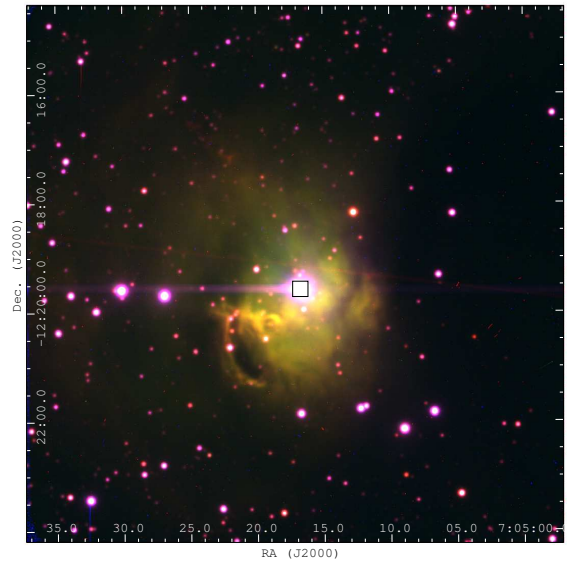}
\caption{The optical color composite image of
Sh2-297 region made using 6724 \textrm{\AA}
[SII] image (red), 6563 \textrm{\AA} H$\alpha$ image (green)
and 5007 \textrm{\AA} [OIII] image (blue), observed using HCT (see
Section \ref{subsection_Optical_photometry}).
The black square marks the position of the ionizing star HD 53623.}
\label{fig_Region}
\end{figure}

Sh2-297 \citep{sha59} is an optically visible (see Figure \ref{fig_Region}), classical Galactic H~{\sc ii} region 
($\alpha_{2000}$ $\sim$ 07$^{h}$05$^{m}$13$^{s}$, $\delta_{2000}$ $\sim$ -12$^{o}$19$^{'}$00$^{''}$), 
ionized by a massive star \object{HD 53623}, and bounded by a cold, dark cloud \object{LDN1657A} to the west. 
It is associated with the reflection nebula \object{IRAS 07029-1215} and
is a part of Canis Major (CMa) OB1 association. In the literature, 
the distance estimates for \object{Sh2-297 range} from 1 to 1.4 kpc \citep{fel81,bic03,for04}. 
In the present study, we have adopted the value of 1.1 kpc from \citet{bic03}. 
\citet{rup66} established the boundaries of 
\object{CMa OB1} as $222^{o}<l<226^{o}$ and $-3.4^{o}<b<+0.7^{o}$. A host of features 
populate the \object{CMa OB1} association - more than 30 nebulae \citep{her78}, 
the \object{CMa R1} association containing a group of stars (including \object{HD 53623}) embedded in a
prominent reflection nebula \citep{van66}, and the three interconnected 
H~{\sc ii} regions of Sh2-292, Sh2-296, and \object{Sh2-297} 
\citep[see][and references therein]{gre08}. \citet{her77}, based on 
the H~{\sc i} morphology \citep{wea74}, emission nebulosity, and the distribution \& properties of 
high-mass OB sources in the region, postulated that a supernova remnant (SNR) could 
have induced star formation in \object{CMa R1}. They estimated the age of the supernova shell to be 
about half a million years. The SNR hypothesis was supported by the work of 
\citet{her78} - who estimated a similar age for most of the 
stellar sources in \object{CMa R1} using optical and infrared studies; and that of 
\citet{vrb87}, whose linear polarization survey of \object{CMa R1} is consistent with the model 
of supernova induced compression of an initially uniform magnetic field, though, 
it must be noted that stellar winds too can produce a shell like structure.
However, an alternate star formation hypothesis by \citet{rey78} - based on the 
study of gas velocities and the fact that UV fluxes from two hot stars 
(HD 54662 and HD 53975) can account for the physical parameters of the shell - 
suggests that strong stellar winds or evolving H~{\sc ii} regions are the cause of star 
formation in the region. \citet{bli80} and \citet{pya86} have 
suggested the same. 

\citet{bon09} have investigated this region using Two Micron All Sky Survey (2MASS) data and concluded that 
it presents similar structural properties as typical young and low mass open 
clusters. \citet{for04} found a deeply embedded,
very young source of intermediate mass, named as UYSO1 (Unidentified
Young Stellar Object 1), that is driving a high-velocity bipolar CO outflow
at the interface of the H~{\sc ii} region and the dark cloud. It was further
resolved into two protostars, UYSO1a and UYSO1b \citep{for09}.
A recent Herschel PACS and SPIRE mapping of the region around these
sources has identified five cool and compact far infrared (FIR) sources,
of masses of the order of a few M$_{\odot}$, hidden in the neighbouring
dark cloud \object{LDN1657A} \citep{lin10}. 

Continuing with our multiwavelength investigations of massive star-forming regions
\citep{ojh04a,ojh04b,ojh04c,sam07,sam10,ojh11}, 
in this paper, we explore the local star formation process around \object{Sh2-297} region.
The presence of multiple facets (like the ionizing star's
radiation field, the nebulosity present in the region, and the dark cloud to the west) necessitates
a multiwavelength study to decipher the star formation scenario. Hence,
we carried out a detailed study of this region in optical, deep NIR,
mid infrared (MIR), and radio wavelengths to learn
about the central ionizing source and its effect on the region, the
physical characteristics of the region, the spatial distribution of
the YSOs and the nature of some selected individual YSOs. Based on 
the study of stellar population, their evolutionary status', and the morphology of the region, 
we have tried to deduce possible star formation activity. 

In Section \ref{section_Obs_and_data_reduction}, we describe our observations and the data reduction procedures.
In Section \ref{section_Other_data_sets}, we have outlined the other data sets which were used
in our study. We discuss the morphology of the region from radio, infrared (IR) and 
submillimetre data in Section \ref{section_Morphology}, and the results pertaining to the stellar 
sources in Section \ref{section_Stellar_population}. In Section \ref{section_StarFormation_Scenario}, 
we discuss the star formation scenario in \object{Sh2-297} region. 
We present our main conclusions in Section \ref{section_Conclusions}. 

\section{Observations and data reduction}
\label{section_Obs_and_data_reduction}

\subsection{Optical photometry}
\label{subsection_Optical_photometry}

Optical photometric observations were performed using the 2 m Himalayan 
\textit{Chandra} Telescope (HCT), Hanle (India), operated
by the Indian Institute of Astrophysics, Bangalore. The telescope 
has Ritchey-Chretien optics with an f/9 Cassegrain focus. 
Both broadband and narrowband observations were performed 
using Himalayan Faint Object Spectrograph and Camera 
(HFOSC) mounted on the telescope. HFOSC has a SITe 
(Scientific Imaging Technologies Inc.) 2k $\times$ 4k pixels CCD, with the central region 
of 2k $\times$ 2k being used for imaging. With a pixel scale of 0.3$^{''}$, this 
translates to a field of view (FoV) of $\sim$ 10$^{'}$ x 10$^{'}$. 
Long and short exposure observations were taken in 
Bessell \textit{V} (600 s \& 20 s) and \textit{I} (300 s \& 10 s) filters 
(centered at $\alpha_{2000}$ $\approx$ 07$^{h}$05$^{m}$13$^{s}$, 
$\delta_{2000}$ $\approx$ -12$^{o}$19$^{'}$03$^{''}$)
on 2009 October 21. [SII] (6724 \textrm{\AA}), H$\alpha$ (6563 \textrm{\AA}), \& [OIII] 
(5007 \textrm{\AA}) filter observations 
(centered at $\alpha_{2000}$ $\approx$ 07$^{h}$05$^{m}$17$^{s}$, 
$\delta_{2000}$ $\approx$ -12$^{o}$19$^{'}$18$^{''}$), with an integration time of 300 s each, 
were done on 2010 January 18. 
Along with the object frames, several bias frames, twilight flat-field frames, and 
the standard field SA98 \citep{lan92} were also observed. 
The average seeing size ranged from $\sim$ 2$^{''}$ to 3$^{''}$ full width half maxima (FWHM).
The standard field was used to 
calculate the extinction coefficients so as to calibrate our CCD systems. 
The astrometric calibration, with a position accuracy better than 0.3$^{''}$, was done 
using the 2MASS sources in the field.  For all the filters, frames of equal 
exposure time were co-added to improve the signal-to-noise
ratio. Photometry was carried out only for \textit{V} and \textit{I} filters. 
Image Reduction and Analysis Facility\footnote{IRAF is distributed by the 
National Optical Astronomy Observatory, which is operated by the Association 
of Universities for Research in Astronomy Inc. under cooperative agreement 
with National Science Foundation.} (IRAF) data reduction package was used 
for the initial CCD image processing. Thereafter, using DAOPHOT II package of 
MIDAS\footnote{MIDAS is developed and maintained by the European Southern Observatory.}, 
the point spread function (PSF) photometry and the photometric measurements were performed. 
A variable PSF was made using several uncontaminated stars in the field. 
Finally, the magnitudes (\textit{V} and \textit{I}) were calibrated in the manner outlined by \citet{ste87}, 
with a precision better than 0.05 mag. 
In the present work, we have used only those sources which have magnitude errors 
$<$ 0.15 mag ($\sim$ 3$\sigma$).  

\subsection{Optical spectroscopy}

The optical spectroscopic observations were carried out using
HCT and the 2 m IUCAA Girawali Observatory (IGO) telescope at Girawali near Pune (India), operated by
Inter University Centre for Astronomy \& Astrophysics (IUCAA), Pune.

\begin{figure}
\centering
\subfigure
{
\includegraphics[width=3.5in,height=3.0in]{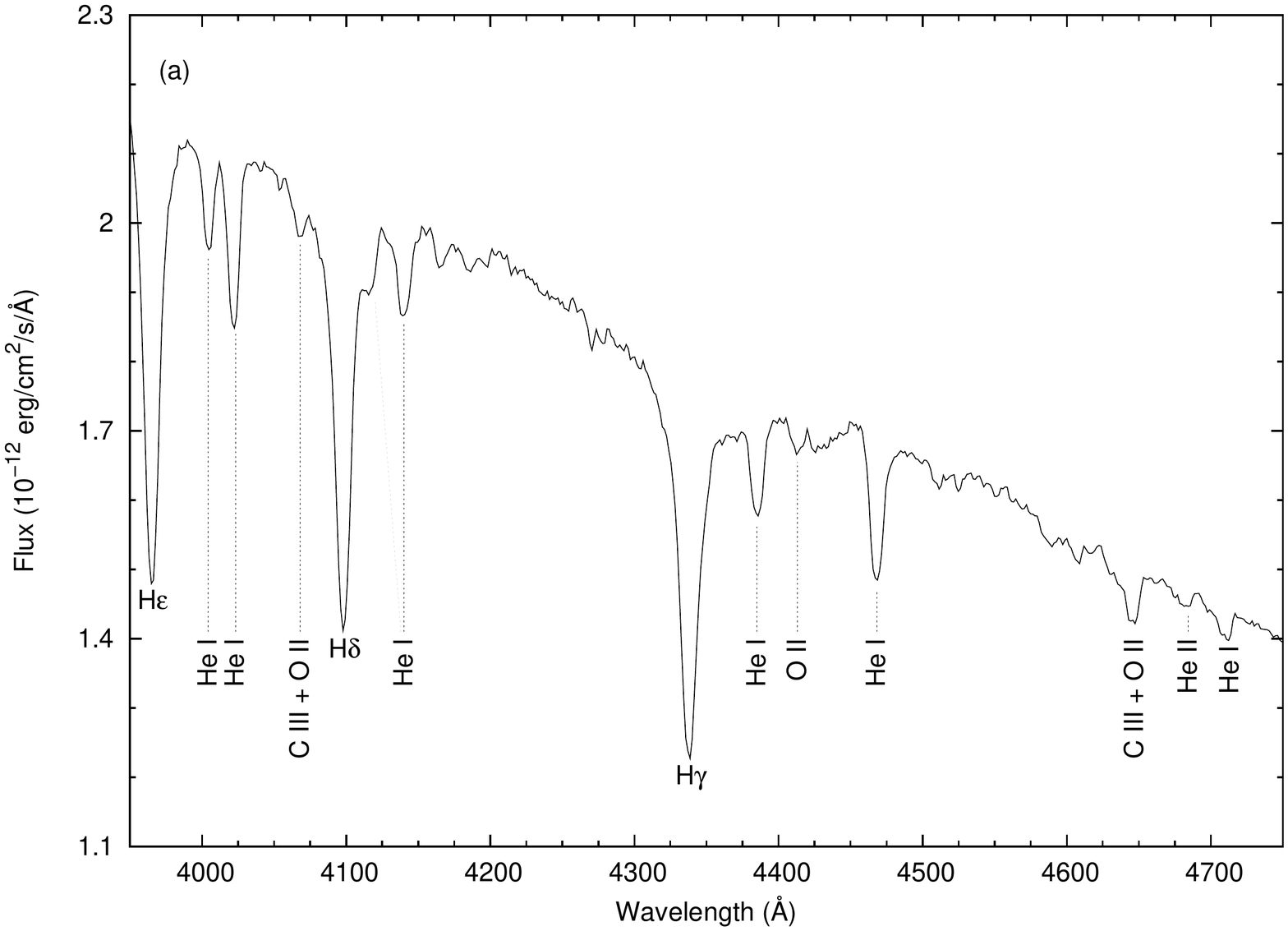}
\label{fig_HCT_spectrum}
}
\subfigure
{
\includegraphics[width=3.5in,height=3.0in]{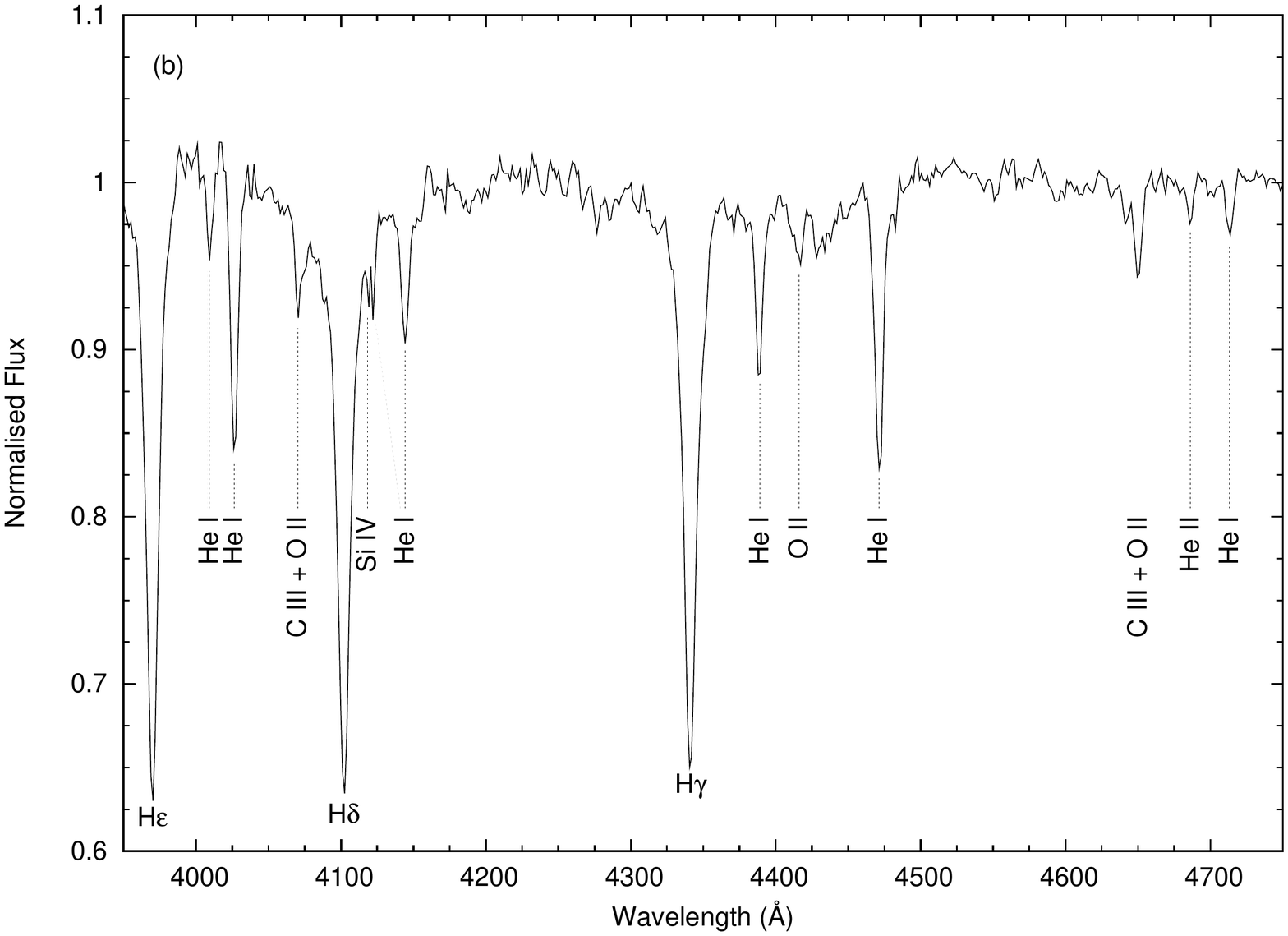}
\label{fig_IGO_spectrum}
}
\caption{(a) Flux-calibrated spectrum of the central ionizing star, HD 53623, obtained with HCT
(b) Normalized spectrum of HD 53623 obtained with IGO.}
\label{fig_spectra}
\end{figure}

The HCT data was obtained on 2009 December 21 using HFOSC, 
with the help of Grism 7 (3500 to 7000 $\textrm{\AA}$) - which has a resolving power of
1200 and a spectral dispersion of 1.45 $\textrm{\AA}$ pixel$^{-1}$. 
The spectrum of the star \object{HD 53623} ($\alpha_{2000}$ $=$ 07$^{h}$05$^{m}$16.75$^{s}$, 
$\delta_{2000}$ $=$ -12$^{o}$19$^{'}$34.5$^{''}$; V=7.99 mag) with an exposure time
of 480 s, in addition to FeAr lamp arc spectrum (for wavelength calibration)
and multiple bias frames, was obtained. The spectrophotometric
standard star G91B2B \citep{mas88} was also observed with an exposure time of 1200 s.
Spectral reduction was carried out using {}``APALL''
task in IRAF reduction package. Finally, the flux-calibrated 
spectrum was obtained (see Figure \ref{fig_HCT_spectrum}). 

To cross-verify the weak diagnostic lines in the object spectrum from HCT, we also obtained a 
deeper spectrum (for 1200 s) for \object{HD 53623} using IGO telescope, on 2009 November 26. 
The telescope has a Cassegrain focus
with a focal ratio of f/10. IUCAA Faint Object Spectrograph and
Camera (IFOSC), with a 2k $\times$ 2k pixel CCD array, was used for the purpose.
Grism 7, with a range of 3800-6840 \textrm{\AA}, resolving power of 1300, 
and a spectral dispersion of 1.49 \textrm{\AA} pixel$^{-1}$ was used. 
Multiple bias frames, halogen flat-field frames, 
and HeNe lamp arc spectrum (for wavelength calibration)
were obtained along with the object spectrum. 
The reduction was done using {}``DOSLIT'' task 
in IRAF. No flux calibration was done for this case. Finally, the
normalized spectrum was obtained (see Figure \ref{fig_IGO_spectrum}). 

To identify the H$\alpha$ emission line stars, two grism
slitless spectra (420 s each) - using Grism 5 (5200-10300 \textrm{\AA}) and H$\alpha$ broadband filter
($\lambda =$ 6563 \textrm{\AA}, $\triangle\lambda =$ 500 \textrm{\AA}) 
- were also obtained with HCT on 2007 November 16 
(centered at $\alpha_{2000}$ $\approx$ 07$^{h}$05$^{m}$23$^{s}$, 
$\delta_{2000}$ $\approx$ -12$^{o}$19$^{'}$00$^{''}$). 
As a reference, for the identification of stars whose slitless spectra were obtained, 
two H$\alpha$ broadband filter images of the field were taken as well. 
Here, the first step was to average the two photometric H$\alpha$ filter
images and the two slitless spectra of the region to improve their
signal-to-noise ratio. The world coordinate system (wcs) coordinates
were then implemented on the averaged photometric image by identifying
a bright set of stars in the FoV, from the USNO catalogue,
and finally synthesizing their J2000 coordinates and pixel coordinates
from the image, using the generic IRAF commands {}``CCMAP'' and {}``CCSETWCS''. 
The averaged spectrum
was then analyzed in ds9 image display device, and the stars with
an H$\alpha$ emission line (detectable as an enhancement over the continuum) 
were identified manually by careful visual inspection, with the help of photometric 
image of the field.
19 H$\alpha$ emission line stars were identified. The detection limit of this identification is $\sim$
3.3 \textrm{\AA} in terms of equivalent width, with the faintest star having a \textit{V} magnitude of 
20.07. Fifteen were located in our NIR FoV, and hence the NIR magnitudes for the remaining 
4 were taken 
from 2MASS catalogue for further analysis.

\subsection{Near infrared observations}
\label{subsection_NIR_observations}

The region of interest, being a star forming region, has considerable 
nebulosity and dark cloud to the west, and hence is obscured at optical wavelengths. 
Moreover, we need to identify and classify the embedded YSOs 
which emit at longer wavelengths, hence we use NIR observations.
NIR observations, centered on
$\alpha_{2000}$ $\approx$ 07$^{h}$05$^{m}$11$^{s}$, 
$\delta_{2000}$ $\approx$ -12$^{o}$18$^{'}$51$^{''}$ for the object,
in $J$ ($\lambda=1.25\,\mu$m), $H$ ($\lambda=1.63\,\mu$m)
and $K_{\mbox{s}}$ ($\lambda=2.14\,\mu$m) bands, were carried out on 2007 February 27 
using the f/10 Cassegrain 1.4 m Infrared Survey Facility (IRSF) telescope, South Africa 
with SIRIUS (Simultaneous InfraRed Imager for Unbiased Survey) - 
a three color simultaneous camera. The camera is equipped with three 1k $\times$
1k HgCdTe arrays, with a FoV of $\sim$ 7.8$^{'}$ $\times$ 7.8$^{'}$, and a pixel scale
of 0.45$^{''}$. Further information about
the instrument can be found in \citet{nag99} and \citet{nag03}.  
For each band, 40 frames each, centered at 9 dithered 
positions, were taken. The integration time for each frame was 
10 s, giving a total integration time of 3600 s (9$\times$40$\times$10) 
in each band for the \object{Sh2-297} region. 
The sky condition was photometric, with the average seeing size ranging from 
1.35$^{''}$ to 1.60$^{''}$ FWHM during the observations.  

The standard data reduction procedure - involving bad pixel masking, 
dark subtraction, flat field correction, sky subtraction, combination 
of dithered frames, adding wcs coordinates to the image and 
PSF photometry - was carried out using IRAF.
Astrometric calibration was implemented using the 2MASS
sources in the field. About 30-40 stars, with good profiles, were selected for calibration for 
each band, and an accuracy of better than $\pm$ 0.04$^{''}$ was achieved 
for our absolute position calibration. 
After the initial processing, photometry was carried out on a final FoV 
of $\sim$ 7.5$^{'} \times$ 7.5$^{'}$. 
The PSF photometry was done using {}``ALLSTAR'' algorithm of 
{}``DAOPHOT'' package in IRAF. Around 11 to 13 isolated bright stars
were taken to make a PSF for each band. Subsequently, the final magnitudes 
were calibrated using the 2MASS stars in the field, with an rms of $\sim$ 0.07 mag
for all the filters. 
To ensure good photometric accuracy, we restricted our catalogue to
sources having uncertainty $<$ 0.15 mag ($\sim$ 2$\sigma$), independent of brightness, 
in all the three bands.
Also, it was found that the sources with 
$K_{\mbox{s}}$ magnitude $\leq$ 8 were saturated in our catalogue; 
hence their magnitudes, in all three filters, were replaced by the 
corresponding 2MASS values. 
We evaluated the completeness limits for our images by doing artificial 
star experiments. Stars of varying magnitudes were added to the images, and then it was determined what 
fraction of stars were recovered in each magnitude bin. The recovery rate was greater than 
90$\%$ for magnitudes brighter than 17, 17, and 16 in $J$, $H$, and $K_{\mbox{s}}$ bands, respectively.
The recovery percentage fell to 85$\%$ for $K_{\mbox{s}}$ magnitudes in 16 to 16.5 mag bin. 
The change in completeness limit is negligible on using the photometric error cutoff at  
0.15 mag. 

\begin{figure}
\includegraphics[scale=0.4]{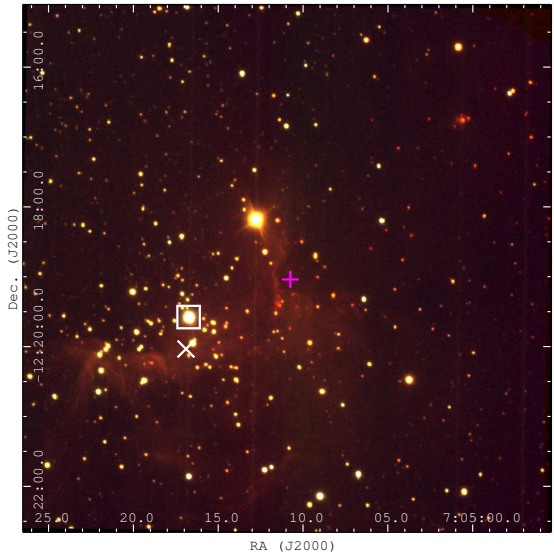}
\caption{Color composite made using \textit{J} (blue), \textit{H} (green) and $K_{\rm s}$ (red)
images. White square marks the position of ionizing star HD 53623, white cross shows the position of
IRAS 07029-1215, and the magenta plus marks the location of UYSO1 from \citet{for04}.}
\label{fig_NIR_ColorComposite}
\end{figure}

The final calibrated catalogue was 
analyzed using the CC and CM diagrams (see Section \ref{section_Stellar_population}). 
For this purpose, the required magnitudes and colors were transformed into 
CIT (California Institute of Technology) system using the color
transformation equations given on the CALTECH 
website\footnote{http://www.astro.caltech.edu/$\sim$jmc/2mass/v3/transformations/}.
In addition to the target region, a control field centered at
$\alpha_{2000}$ $\approx$ 07$^{h}$06$^{m}$20$^{s}$, 
$\delta_{2000}$ $\approx$ -12$^{o}$01$^{'}$40$^{''}$ 
(about 24$^{'}$ to the north-east of our object field) 
was also observed. It was reduced in a similar manner as the object frames.
The control field was used to assess the level of contamination from foreground 
sources in the target field. It was used to delineate the maximum $H-K$ color for 
the foreground sources, so as to identify the young stellar sources in the region - 
which will have a larger color by virtue of extinction and the presence  
of circumstellar disk which results in excess \textit{K}-band emission. 
Figure \ref{fig_NIR_ColorComposite} shows a false NIR color composite 
(\textit{J} : blue, \textit{H} : green, $K_{\mbox{s}}$ : red) image of Sh2-297 region,
which reveals nebulosity around the central and south-eastern region, 
and the dark cloud towards the west. 

\subsection{Radio continuum observations}

1280 MHz observation was carried out on 2007 July 12,
while that for 610 MHz on 2007 July 19, using Giant
Metrewave Radio Telescope (GMRT) array. The GMRT array consists of
30 antennae, arranged in an approximately Y-shaped configuration,
with each dish being 45 m in diameter. 12 antennae are located randomly
in a compact 1 km$\times$1 km central square area, while the remaining
18 are located along the three radial arms (6 along each arm), each
arm length being $\sim$ 14 km. The maximum baseline is $\sim$ 25 km.
Further details can be found in \citet{swa91}. For our 
observations, VLA phase calibrator 
{}``0735-175'' and flux calibrator {}``3C147'' were used,
at both the frequencies. The details of the observations are given in Table \ref{table_Radio_data}. 

\begin{deluxetable*}{c c c}
\tablecolumns{3}
\tablewidth{0pt}
\tabletypesize{\scriptsize}
\tablecaption{Details of GMRT Observations}
\tablehead{
\colhead{} & \colhead{1280 MHz} & \colhead{610 MHz}}
\startdata
Date of Obs. & 2007 July 12 & 2007 July 19 \\
Phase Center & $\alpha_{2000}$ = 07$^h$05$^m$12.30$^s$ & $\alpha_{2000}$ = 07$^h$05$^m$12.30$^s$ \\
             & $\delta_{2000}$ = -12$^o$19$^{'}$01.14$^{''}$ & $\delta_{2000}$ = -12$^o$19$^{'}$01.14$^{''}$ \\
Flux Calibrator & 3C147 & 3C147 \\
Phase Calibrator & 0735-175 & 0735-175 \\
Cont. Bandwidth & 16 MHz & 16 MHz \\
Primary Beam & 24$^{'}$ & 43$^{'}$ \\
Resolution of maps  &  & \\
used for fitting & 23.5$^{''}\times$13.9$^{''}$ &  23.8$^{''}\times$13.6$^{''}$ \\
Peak Flux Density & 19.42 mJy/beam & 41.33 mJy/beam \\
rms noise & 2.32 mJy/beam & 3.24 mJy/beam \\
Integrated Flux Density & 0.67 Jy & 0.64 Jy \\
\enddata
\label{table_Radio_data}
\end{deluxetable*}

The radio data from GMRT was analyzed with the help of Astronomical Image
Processing Software (AIPS) distributed by NRAO. 
The data was edited to flag the 
bad antenna pairs, baselines, and time ranges using the 
{}``VPLOT'', {}``UVFLG'', and {}``TVFLG'' tasks. 
The final image was produced using a set of tasks which involved fourier inversion,
cleaning (using the task {}``IMAGR'' in AIPS), followed by a
few iterations of phase-only self-calibration (till the rms convergence) to remove
the effects of the atmospheric and ionospheric phase distortions.

Also, since our source is located near the Galactic plane, we had
to take into account the system temperature corrections for GMRT.
Since, at metre wavelengths, there is a large amount of radiation from the Galactic 
plane, the effective temperature of the antennae increases, 
which has to be corrected for. 
This was done by obtaining the sky temperature map of \citet{has82} 
at 408 MHz, and rescaling the image by a scaling factor which is just the ratio of
system temperature towards the target source and the flux calibrator (here 3C147). 
The temperature for the source will be T$_{\mbox{f}}$ + T$_{\mbox{sys}}$, 
where T$_{\mbox{f}}$ = T$_{408}$ $\times$(f/408)$^{-2.6}$ \citep[using the spectral index given in][]{has82},
with T$_{408}$ being the temperature value from Haslam's map at 408 MHz and {}``f'' being
the frequency (in MHz) for which the correction is to be calculated.
T$_{\mbox{sys}}$, the system temperature, was obtained from the National
Centre for Radio Astrophysics (NCRA) website\footnote{http://www.ncra.tifr.res.in/$\sim$gtac/GMRT-specs.pdf} 
for both the frequencies. The temperature for the flux calibrator will merely be T$_{\mbox{sys}}$,
hence the scaling factor is (T$_{\mbox{f}}$ + T$_{\mbox{sys}}$)/T$_{\mbox{sys}}$.

In addition, for 1420 MHz, the
archival image from Very Large Array (VLA), for the observation
date 1998 December 14, was also used (Project ID AF346). 

Radio continuum interferometric data, at 610 MHz, 1280 MHz and 1420
MHz was used to probe the ionized gas component.
The GMRT maps were convolved to a 
resolution of about 24$^{''} \times$14$^{''}$, approximately same as the VLA map. After obtaining the contour
maps for all the three frequencies, the integrated flux was found
by fitting a 3$\sigma$ contour level box and using the AIPS
task {}``IMSTAT'' (see Table \ref{table_Radio_data}). Using these radio continuum data points, the various
physical parameters of the region were calculated (see Section \ref{subsection_Physical_properties}).

\section{Other available data sets}
\label{section_Other_data_sets}

\subsection{Near infrared data from 2MASS}

We used the 2MASS\footnote{This publication makes use of 
data products from the Two Micron All Sky Survey, which is a joint project of the 
University of Massachusetts and the Infrared Processing and Analysis Center, California 
Institute of Technology, funded by NASA and the NSF.} 
data in the field for our photometric and astrometric calibrations.
The stars were obtained using the task {}``IMTMC'' in {}``WCSTOOLS''
package in IRAF, for the $J$ ($\lambda=1.25\,\mu$m), $H$ ($\lambda=1.63\,\mu$m)
and $K_{\mbox{s}}$ ($\lambda=2.14\,\mu$m) bands. All the sources which
were obtained for calibration had a good quality flag value of {}``AAA'' for
the magnitudes. We also substituted the 2MASS magnitudes for those sources which were 
found to be saturated in our IRSF catalogue, i.e. those 
for which we had IRSF $K_{\mbox{s}}$ magnitude $\leq$ 8.

\subsection{Near and mid infrared data from WISE}

Wide-field Infrared Survey Explorer (WISE) surveyed the entire
sky in the NIR bands at 3.4 $\mu$m (w1) and 4.6 $\mu$m (w2), and in the MIR bands at
12 $\mu$m (w3) and 22 $\mu$m (w4), with an astrometric precision better 
than $\sim$ 0.15$^{''}$ \citep{wri10}. 
The spatial resolution was  $\sim$ 6$^{''}$ for the first three
bands and $\sim$ 12$^{''}$ for 22 $\mu$m. 

Since the WISE (Preliminary Release April 2011) catalogue had many sources missing
(e.g. for w1 band, only $\sim$ 35$\%$ of the sources detected in our photometry were 
present in the release catalogue), as could also
be visually discerned, we carried out aperture 
photometry on the archival WISE images - for the same FoV as our IRSF NIR field 
- following the method described by \citet{koe12}. This significantly 
improved the number of sources detected. 
Photometry was done using the DAOPHOT package in IRAF; with the 
aperture radius, inner radius of the sky annulus, \& outer radius of the 
sky annulus being 5$^{''}$, 7.5$^{''}$, \& 25$^{''}$ for the 3.4 \& 4.6 $\mu$m bands, 
and 7.5$^{''}$, 10$^{''}$, \& 25$^{''}$ for the 12 $\mu$m band, respectively. 
The 22 $\mu$m band was not used mainly because of its low resolution, 
as well as the dominant nebulosity and the presence of not many point sources 
(12 sources were detected out of which 9 were in highly nebulous central region). 
The photometry for each band was 
calibrated with the WISE preliminary release catalogue. Care was taken not 
to use the objects flagged as {}``D'' (diffraction spikes), {}``H'' (halos from bright sources),
{}``P'' (persistent artifacts), or {}``O'' (optical ghosts of bright sources) in 
the release catalogue for calibration. 
For the final WISE catalogue, we took only those sources which had 
all magnitude errors $\textless$ 0.3, to make sure that they had reliable photometry.

Our IRSF NIR catalogue was subsequently matched with the WISE catalogues 
within a 3$^{''}$ matching radius. 
187, 124, and 9 sources detected in our NIR FoV 
had 3.4 $\mu$m, 4.6 $\mu$m, and 12 $\mu$m counterparts, respectively. 
The resultant final catalogue was used to get the SEDs 
of various sources, and hence in constraining the manifold physical 
parameters like age, mass, accretion rates etc. for them (see Section \ref{section_SED}). 
The 3.4 $\mu$m and 12 $\mu$m bands contain prominent 
polycyclic aromatic hydrocarbon (PAH) features at 3.3, 11.3 \& 12.7 $\mu$m \citep{wri10,sam07}
in addition to the continuum emission, and
hence can be used to get an idea of the photodissociation region. 
The 22 $\mu$m band can be used to examine the warm dust emission - the stochastic emission from 
small grains as well as the thermal emission from large grains \citep{wri10}.

\section{Morphology of the region}
\label{section_Morphology}

\subsection{The ionizing source}

The ionizing source of \object{Sh2-297} region is the star \object{HD 53623}, associated with the 
reflection nebula \object{CMa R1} \citep{van66}. 
\citet{cla74} and \citet{hou88} estimated the spectral type of this source 
as B1Vn and B1II/III, respectively, using objective-prism plates from Michigan Curtis-Schmidt
telescope, CTIO. \citet{her78} have proposed the spectral type as B0.5 IV-V using spectrogram analysis. 
Here we use the optical spectroscopic 
data to find out the spectral type of this source. 
The reduced spectra from HCT and IGO, for the wavelength range 
3950 to 4750 $\textrm{\AA}$, have been shown in Figure \ref{fig_spectra}. 
The spectra were analyzed with the help of Walborn
spectra catalogue \citep{wal90}.
The absence of He~{\sc ii} lines at 4200 \& 4541 $\textrm{\AA}$, and 
low strength at 4686 $\textrm{\AA}$  
suggest that the star does not belong to O-type spectral class. 
The most prominent lines in our spectra
are the He~{\sc i} lines at 4009, 4026, 4144, 4387, 4471, and 4713 $\textrm{\AA}$.
The Balmer lines at 3970 $\textrm{\AA}$ (H$\varepsilon$), 4102 $\textrm{\AA}$ (H$\delta$),
and 4341 $\textrm{\AA}$ (H$\gamma$) are clearly seen too. 
We also see C~{\sc iii} + O~{\sc ii} blend at 4070 and 4650 $\textrm{\AA}$. 
Weak Si~{\sc iv} line at 4116 $\textrm{\AA}$, weak O~{\sc ii} lines at 4415-4417 $\textrm{\AA}$, 
and absence of Si~{\sc iii} lines indicate that the star most probably 
does not belong to giant luminosity classes.
The preponderance of neutral helium lines and the comparison with the 
standard Walborn catalogue suggest a most likely spectral type of B0V/B0.5V for the star.
\citet{tji01} have suggested, using H$\alpha$ emission study, that this 
source does not contain any circumstellar disk, most probably due to photoevaporation
by UV photons or due to influence of some supernova in the nearby region.

\subsection{Physical properties of the region}
\label{subsection_Physical_properties}

\begin{figure}
\centering
\subfigure
{
\includegraphics[scale=0.35]{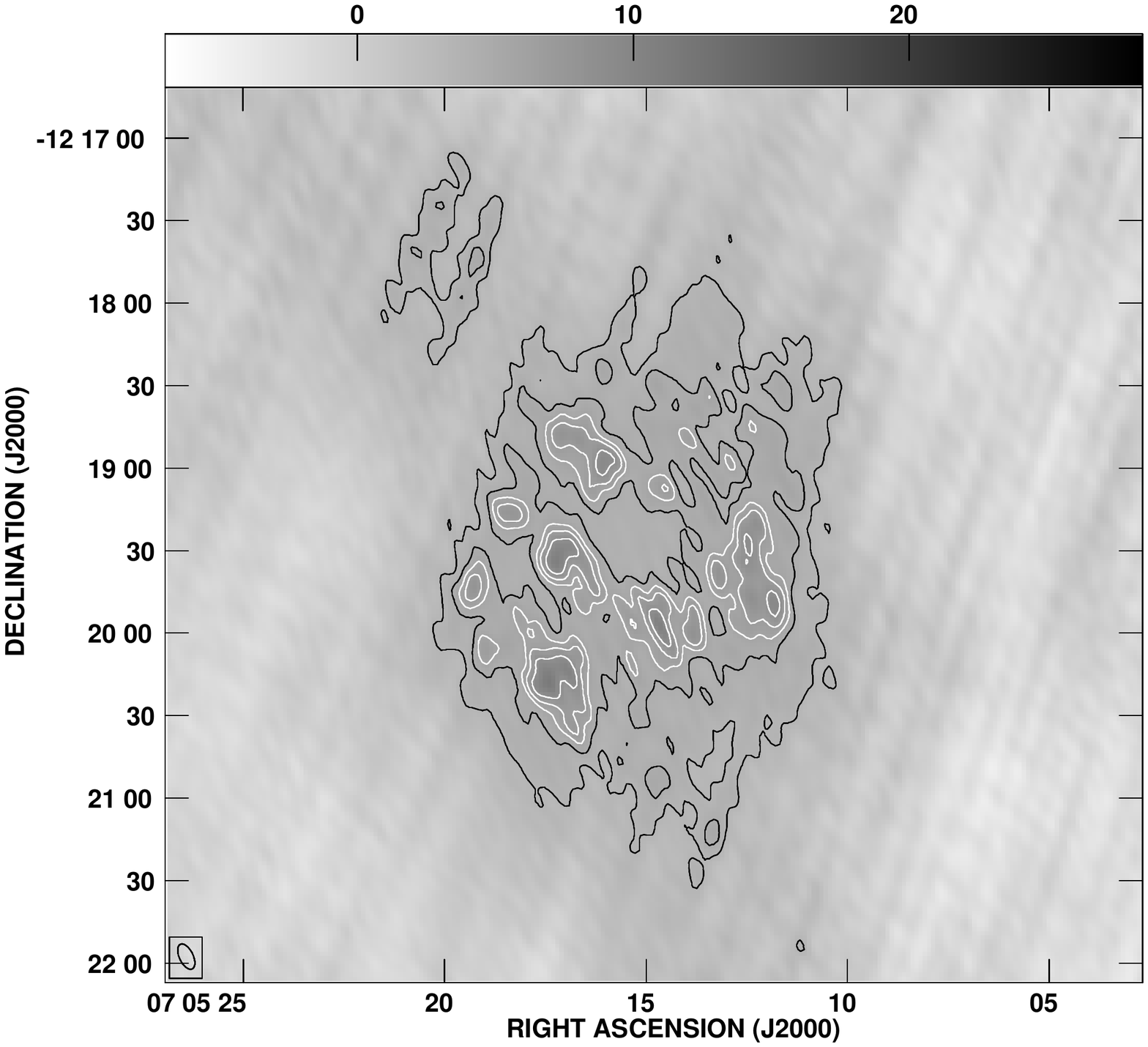}
}
\subfigure
{
\includegraphics[scale=0.35]{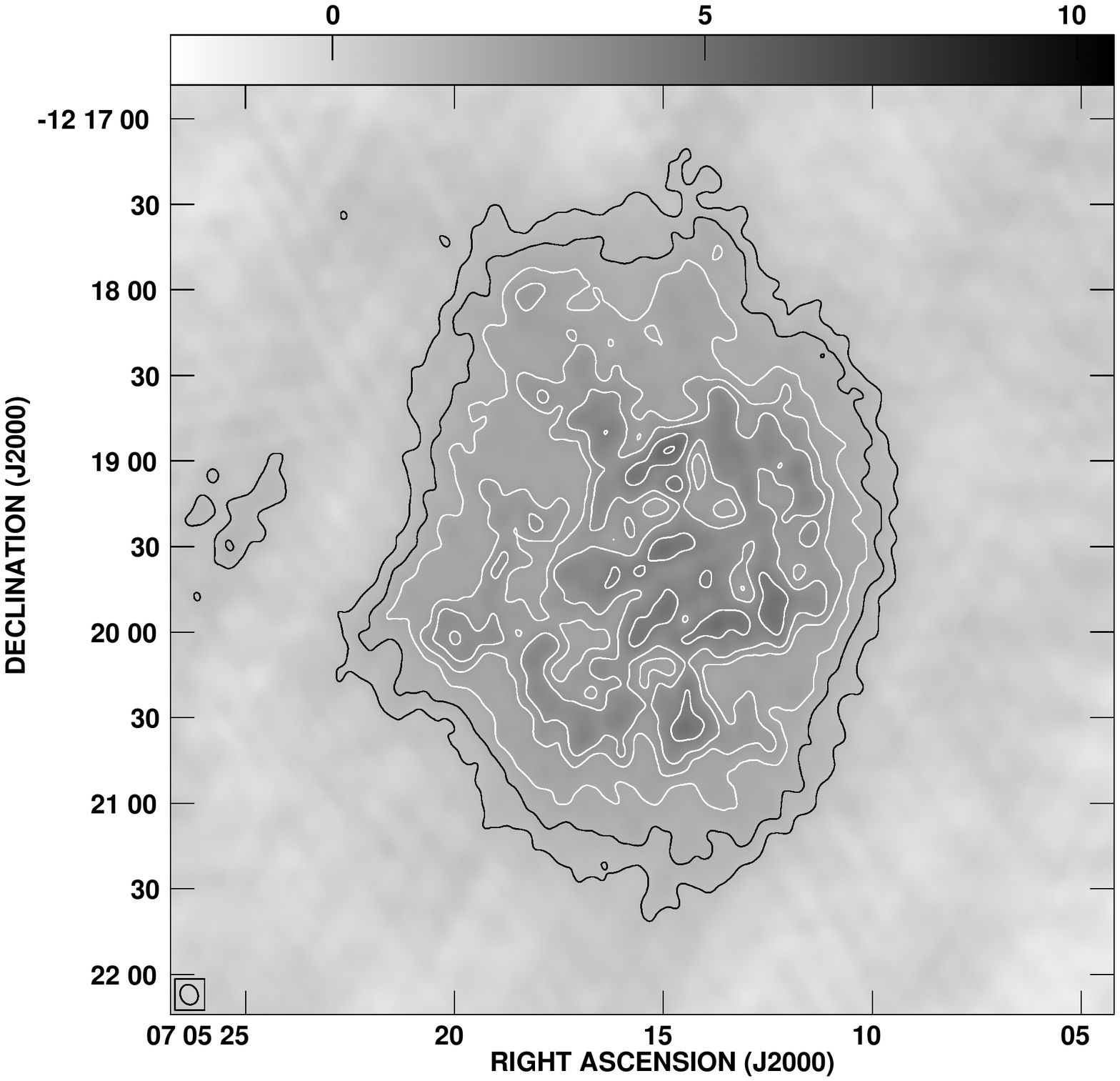}
}
\caption{($upper\,panel$) High resolution GMRT radio map for 610 MHz ($\sim$ 10$^{''}\times$5$^{''}$).
Contours have been drawn at 12, 16, 20, 24, \& 30 \% of the peak flux (28.30 mJy/beam).
($lower\,panel$) High resolution GMRT map for 1280 MHz ($\sim$ 7$^{''}\times$ 6$^{''}$).
Contours have been drawn at 10, 15, 20, 25, 30, 40, \& 50 \% of the peak flux (10.47 mJy/beam).}
\label{fig_Radio_highresolution}
\end{figure}

High resolution studies are important as they help us in understanding the 
structure of a region. \object{Sh2-297} has been observed with very coarse resolutions 
in previous radio surveys, which makes it harder to recognize the subtler features 
in the radio emission. For example, the survey by \citet{fel72}  
observed 168 optically identified H~{\sc ii} regions at 1400 MHz, including \object{Sh2-297}, with a spatial resolution 
of 10$^{'}$. Similar observations were carried out by \citet{pya80} for 
wavelengths ranging from 2 to 11 cm, with resolutions of $\sim$ 1$^{'}-$4.3$^{'}$. 
At larger frequencies (10 GHz), the CMa R1 region was observed by \citet{nak84} with a 
resolution of 2.7$^{'}$. 
\citet{fic93}, in a VLA survey, observed the region at 1.4 GHz with 43$^{''}$ resolution. 
Figure \ref{fig_Radio_highresolution} shows GMRT high resolution greyscale contour maps, 
for 610 MHz ($\sim$ 10$^{''} \times$ 5$^{''}$) and 1280 MHz ($\sim$ 7$^{''} \times$ 6$^{''}$),
of \object{Sh2-297} H~{\sc ii} region. The intricate morphology of \object{Sh2-297} - showing the complex structures 
and the clumpy nature - can very well be discerned on examination. 
The radio emission morphology from our observation as well as the location of H~{\sc ii} region 
at the tip of the molecular cloud (see Section \ref{subsection_IR_and_submm_structures}) suggest that the ionized 
gas is possibly undergoing a champagne flow (also known as blister region) \citep{ten79,all79}, with the extended 
emission towards the east and the head of the ionized champagne flow located 
to the west towards \object{LDN1657A} (see Section \ref{section_StarFormation_Scenario}). 

\begin{figure}
\vspace*{-16.5cm}
\includegraphics[scale=0.8]{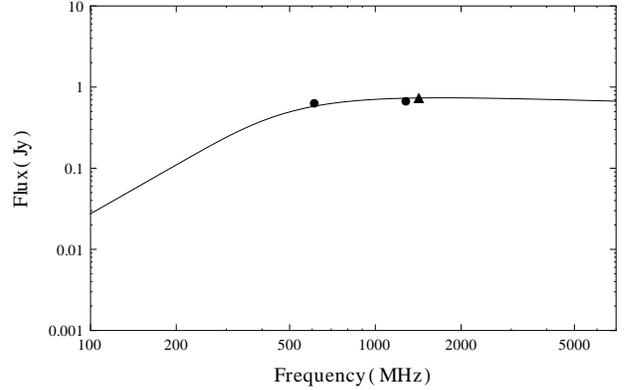}
\caption{Observed and fitted flux density (solid line) for the H~{\sc ii} region.
Filled circles and triangle represent GMRT \& VLA data, respectively.
The flux densities were calculated for an approximately same
resolution of $\sim$ 24$^{''}\times$14$^{''}$.}
\label{fig_Radiofit}
\end{figure}

The radio continuum observations of the region using GMRT, and archival
data from VLA, were used to calculate the various physical parameters
which characterize an H~{\sc ii} region (emission measure, electron density, 
str\"{o}mgren radius, and dynamical timescale). 
Assuming a spherically symmetric and homogeneous ionized region, the model of 
\citet{mez67} for free-free emission was used. 
According to this, the flux density arising due to the region 
is given by the following \citep[adapted from][Equations 1 and 3]{mezEtAl67} :

\begin{equation}
S_{\nu}=3.07\times10^{-2}T_{e}\nu^{2}\Omega(1-e^{-\tau(\nu)})
\end{equation}

\begin{equation}
\tau(\nu)=1.643\times10^{5}aT_{e}^{-1.35}\nu^{-2.1}n_{e}^{2}l
\end{equation}

where, $S_{\nu}$ is the integrated flux density in Jansky (Jy), $T_{e}$
is the electron temperature of the ionized core in K, $\nu$ is the
frequency in MHz, $n_{e}$ is the electron density in cm$^{-3}$,
$l$ is the extent of the ionized region in pc, $\tau$ is the optical
depth, $a$ is the correction factor, and $\Omega$ is the solid angle
subtended by the beam in steradian (approximating the beam with a Gaussian,
the beam solid angle is given by $\Omega=1.133 \times \theta_{maj} 
\times \theta_{min}$, where $\theta_{maj}$ and $\theta_{min}$ are 
major and minor half power beam widths). The factor $n_{e}^{2}l$
denotes the emission measure (in cm$^{-6}$ pc), a measure of optical
depth of the medium. For our calculations, we set the value of $a$
as 0.99 \citep[obtained from][Table 6]{mez67}, and that of $T_{e}$ to 10000 K 
- applicable to classical H~{\sc ii} regions \citep{sta04}. Thereafter, 
treating the emission measure as a free parameter
and using the three data points (flux density vs. frequency) from
the three maps, a non-linear regression was implemented,
following a procedure similar to \citet{vig06} (see Figure \ref{fig_Radiofit}), which yielded
the value of the free parameter (emission measure), for the best fit, as 9.15 $\pm$
0.56 $\times$ 10$^{5}$ cm$^{-6}$ pc. The extent of this extended H~{\sc ii}
region is $\sim$ 5$^{'}$, as obtained from the VLA image. At a distance of
1.1 kpc, this translates to an extent of 1.6 pc. From the emission
measure and the extent, the electron density is obtained to be 756
$\pm$ 46 cm$^{-3}$. The low values of the emission measure and the
electron density reinforce the fact that this is a classical,
and thus, evolved H~{\sc ii} region  \citep[see][]{sta04}. 

Further, assuming a homogeneous and spherically symmetric nature for
the H~{\sc ii} region, the luminosity of the Lyman continuum photons (in photons s$^{-1}$) from
the central ionizing source is given by \citep[Equation 5]{mor83} :

\begin{equation}
S_{*}=8\times10^{43}\left(\frac{S_{\nu}}{mJy}\right)\left(\frac{T_{e}}{10^{4}K}\right)^{-0.45}
\left(\frac{D}{kpc}\right)^{2}\left(\frac{\nu}{GHz}\right)^{0.1}
\end{equation}

where $S_{\nu}$ is the integrated flux density in mJy from the contour
map, $D$ is the distance in kiloparsec, $T_{e}$ is the electron 
temperature, and $\nu$ is the frequency in GHz for which the luminosity
is to be calculated. For our calculations, we used the 1420 MHz 
flux value (0.76 Jy). $D$ was taken to 
be 1.1 kpc and $T_{e}$ as 10$^{4}$ K. The value for $S_{*}$ was determined
to be 7.62$\times$10$^{46}$ photons s$^{-1}$ (hence $\log S_{*}$ is 46.88). 
\citet{bon09} have estimated the main sequence (MS) age 
of \object{HD 53623} as 5 $\pm$ 3 Myr, therefore
assuming a luminosity class V for it, 
a comparison of $\log S_{*}$ with \citet[Table II]{pan73} 
yields the spectral type as B0, in agreement with our optical spectroscopic analysis. 
The lower value of our calculated luminosity ($\log S_{*}$) as compared to the theoretical 
value (47.63) from \citet{pan73} could be because of absorption of photons
by dust in the region.

As the central massive star ionizes the ambient medium around it, the
ionization front moves outwards, and the whole H~{\sc ii} region expands
till an equilibrium is reached between the number of ionizations and
recombinations. The underlying assumption here is that the ambient
medium has homogeneous density and temperature. Using the formulation
given by \citet{str39}, the radius of the H~{\sc ii} region 
at this point - called Str\"{o}mgren radius - is given by : 

\begin{equation}
R_{s}=\left(\frac{3S_{Lyman}}{4\pi n_{o}^{2}\beta_{2}}\right)^{1/3}
\end{equation}

where $R_{s}$ is the Str\"{o}mgren radius
(in cm), $n_{o}$ is the initial ambient density (in cm$^{-3}$), 
and $\beta_{2}$ is the total recombination coefficient
to the first excited state of hydrogen. The value of $\beta_{2}$, for a $T_{e}$
of 10$^{4}$ K, is taken to be 2.6$\times$10$^{-13}$ cm$^{3}$ s$^{-1}$
\citep{sta04}. 
We estimated the initial ambient density as follows \citep[similar to][]{str39}. 
From the luminosity calculations for Lyman continuum photons, 
we have obtained a calculated value of 7.62$\times$10$^{46}$ photons s$^{-1}$ ($S_{*}$), 
which is about 18\% of the theoretical output of 
4.27$\times$10$^{47}$ photons s$^{-1}$ \citep{pan73}. 
This implies that $n_{e}=0.18 n_{o}$, 
hence yielding $n_{o}$ as 4202 $\pm$ 253 cm$^{-3}$, using the above calculated 
value of the electron density. 
Using these values for $S_{*}$ and $n_{o}$, the str\"{o}mgren radius turns 
out to be 0.051 $\pm$ 0.002 pc.  

As the gas expands, it sends a shock front into
the neutral medium. This shock front precedes the ionization front.
At this stage, the radius of the H~{\sc ii} region is given by \citep{spi78}: 

\begin{equation}
R(t)=R_{s}\left(1+\frac{7c_{II}t}{4R_{s}}\right)^{4/7}
\end{equation}

where $R(t)$ is the extent of the H~{\sc ii} region at time $t$, and
$c_{II}$ is the speed of sound, taken to be 11$\times$10$^{5}$
cm s$^{-1}$ \citep{sta04}. 
Using the current extent of 1.6 pc as the value of $R(t)$, and the above 
calculated value of $R_{s}$ in this equation, we obtain the value of
$t$, referred to as the dynamical timescale, as 1.07 $\pm$ 0.31 Myr.

It should be noted that, to calculate $R_{s}$ and $t$, we have estimated 
$n_{o}$ (4202 $\pm$ 253 cm$^{-3}$) assuming a luminosity class V for the massive star, and that 
only part of the radio flux is used for ionization. Also, we have 
used the current value of electron density. However, at the beginning of massive 
star formation, $n_o$ could be of the order of 10$^4$-10$^5$ cm$^{-3}$ \citep{war11}. 
Besides, the expansion velocity decreases with time. Hence, the electron density 
and the dynamical timescale calculated here should be treated as conservative lower 
limits.

\subsection{Infrared and submillimetre structures}
\label{subsection_IR_and_submm_structures}

\begin{figure}
\includegraphics[scale=0.4]{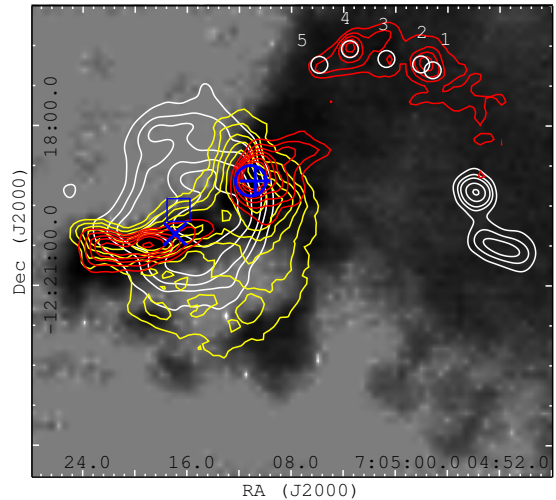}
\caption{The $^{12}$CO(J=3-2) image (log greyscale with intensity range
-1 to 30 K km s$^{-1}$) of the region from JCMT, overlaid with
MSX 8.28 $\mu$m contours (yellow) showing the PAH distribution,
SCUBA 850 $\mu$m contours (red) which trace the cold dust,
and 1280 MHz radio continuum contours (white) indicating the ionized gas. The CO image reveals a
maximum column density of 87.28 K km s$^{-1}$.
MSX and SCUBA contours have been plotted at 10, 15, 20, 25, 30, 40 \& 50 \% of their peak values,
which are 49.666 mW/m$^2$-Sr and 1.619 Jy/beam, respectively.
Radio contours ($\sim$ 25$^{''}$ resolution) have been plotted at
3, 6, 9, 15, and 21$\sigma$ ($\sigma$ denotes the rms noise, which is 1.03 mJy/beam) levels.
The blue square marks the ionizing star HD 53623, blue plus symbol marks
the position of UYSO1 from \citet{for04}, blue circle marks the position of C$_{2}$H peak from \citet{beu08},
blue cross denotes IRAS 07029-1215, and the sources from \citet{lin10}
have been labelled and marked in white circles.}
\label{fig_CO3to2}
\end{figure}

Figure \ref{fig_CO3to2} shows the $^{12}$CO(J=3-2)  
image of the region  
(public processed data for observation date 2011 February 20, Project ID: M11AH48A)\footnote
{The James Clerk Maxwell Telescope is operated by the Joint Astronomy Centre on behalf of the Science 
and Technology Facilities Council of the United Kingdom, the Netherlands Organisation for Scientific 
Research, and the National Research Council of Canada. The archival data was downloaded from 
http://www.cadc.hia.nrc.gc.ca/jcmt/search/product/.} from James Clerk Maxwell Telescope (JCMT). 
The MSX 8.28 $\mu$m 
contours (in yellow), SCUBA 850 $\mu$m contours (in red) \citep{fra08}, and the 1280 MHz 
radio contours from GMRT (in white) are overlaid. 
The MSX\footnote{This research made use of data products from \textit{Midcourse Space Experiment}.
Processing of the data was funded by Ballistic Missile Defense Organization 
with additional support from the NASA Office of Space Science.}
A band ($\lambda =$ 8.28 $\mu$m; containing PAH features at 7.7 \& 8.7 $\mu$m besides continuum emission) 
contours give an idea of the PAH distribution in the region \citep[see e.g.][]{deh05}. 
$^{12}$CO(J=3-2) traces the molecular hydrogen, 
while SCUBA 850 $\mu$m traces the cold dust. 
As can be seen from the figure, there is a clear correlation 
between the ionized gas and 850 $\mu$m contours in Sh2-297 central region. 
The $^{12}$CO image shows a D-shaped ring to the south of \object{HD 53623}, which is also 
traced by the MSX A band contours, and is faintly visible in narrowband H$_{2}$ S(1) image of 
\citet[see their Figure 1]{for09}. This is probably due to the sweeping up and piling of material from the expanding 
H~{\sc ii} region, and as such it is a potential site for future triggered star formation by collect-and-collapse
process. 
Also noticeable is the fact that, towards the west, 
PAH extends roughly upto the interface of the H~{\sc ii} region and LDN1657A.

The WISE w4 band which traces the warm dust and thermal emission from protostars 
(not shown here), and SCUBA data show that the warm and cold dust extend well 
beyond the H~{\sc ii} region, and also towards the cold, dark cloud \object{LDN1657A}. 
The gas and dust correlation is evident in the FIR IRAS 100 
$\mu$m wavelength as well \citep{gre08}. The nebula around the star 
\object{HD 53623} shows a high level of FIR emission.
\citet{beu08} found a C$_{2}$H peak near \object{IRAS 07029-1215}, 
coinciding with UYSO1 (see Figure \ref{fig_CO3to2}),
thereby indicating that this is a young star forming region \citep{vas11}. 
\citet{kim04} have probed 
the dense molecular gas using the optically thin $J=1-0$ transition of $^{13}$CO. 
One of the peaks is at $l=225.47^o$, $b=-2.80^o$ (outside the range of 
Figure \ref{fig_CO3to2}), showing the presence of dense 
gas further towards the south-west of \object{Sh2-297} as well.

\section{Stellar population in the region}
\label{section_Stellar_population}

\subsection{NIR color-color diagram}

\begin{figure}
\centering
\subfigure
{
\includegraphics[width=3.1in]{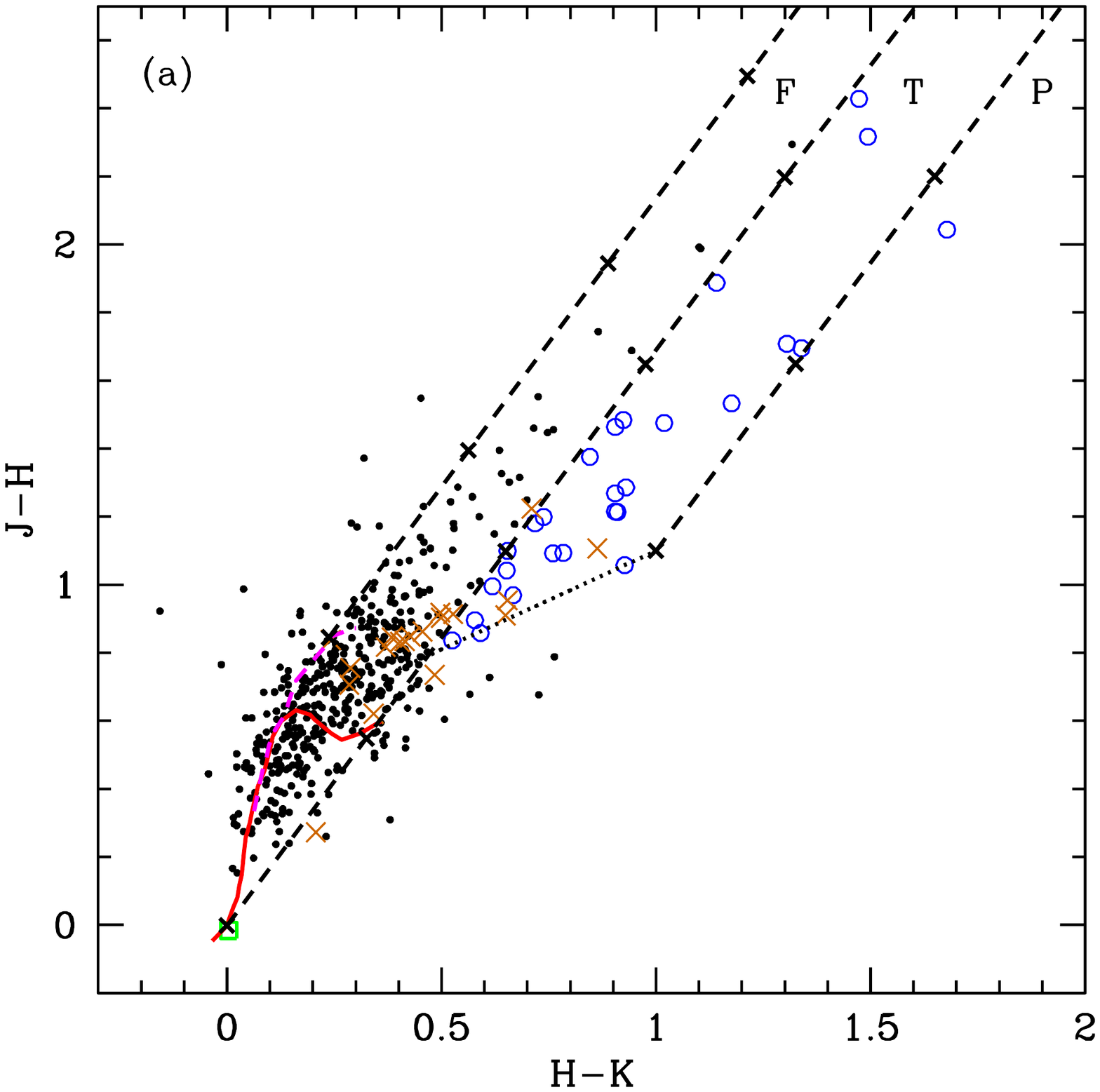}
\label{fig_NIR_CCD_object}
}
\subfigure
{
\includegraphics[width=3.1in]{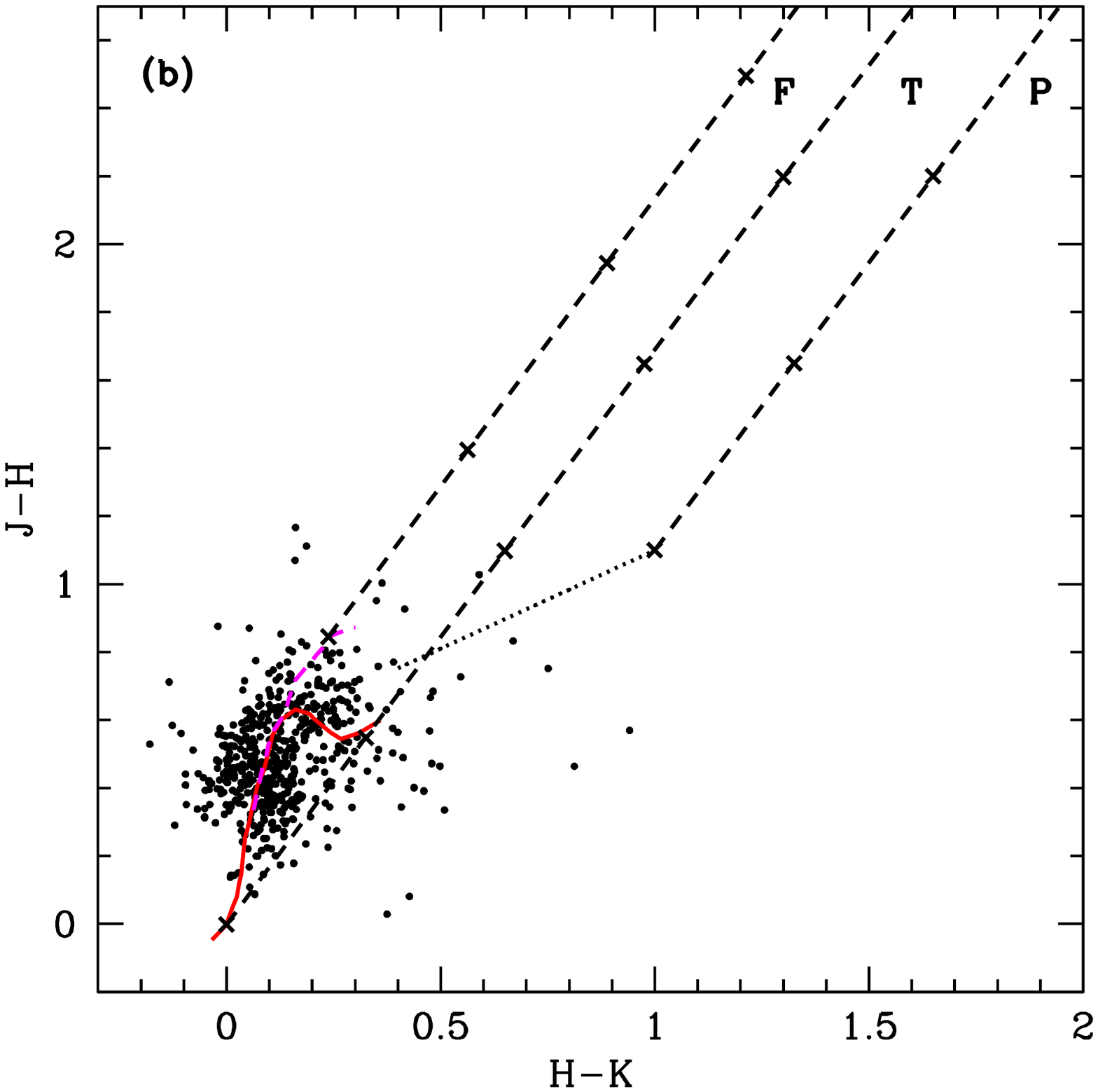}
\label{fig_NIR_CCD_control}
}
\caption{(a) NIR CC diagram for the object field. The red curve shows the MS locus and the magenta
curve shows the locus of giant stars. The locus of CTTS has been shown by a black dotted line.
The three parallel and slanted lines are the reddening vectors, with crosses on them marking an interval of $A_{V}$= 5 mag.
Blue circles denote the Class II-type sources, orange crosses the H$\alpha$ emission line stars,
and the green square the central ionizing star. (b) The CC diagram for the control field.}
\label{nir_CCD}
\end{figure}

The NIR CC diagram for the IRSF sources, showing the $J-H$ vs.
$H-K$ colors, is given in Figure \ref{fig_NIR_CCD_object}. The thick red line denotes
the locus of MS stars, the magenta line shows the locus
of giant stars while the dotted black straight line is the locus of
Classical T Tauri Stars (CTTS). The loci for the MS and
giants were taken from \citet{bes88} while those for the
CTTS from \citet{mey97}. The three parallel dashed and slanted 
lines are the reddening vectors ($A_{J}/A_{V}=0.265$, $A_{H}/A_{V}=0.155$
and $A_{K}/A_{V}=0.090$) for the CIT system from \citet{coh81}. 
The crosses mark every 5 mag increase in $A_{V}$. 
We converted the colors from 2MASS to CIT system,
as was discussed in Section \ref{subsection_NIR_observations}.  

We used the CC diagram to identify and study the nature of YSOs in the region. 
Since young sources have a significant infrared excess,
which in turn can be used to identify them, we divided our CC diagram
into three regions - F, T and P \citep[similar to][]{ojh04b,ojh04c}. 
The region {}``F'' mostly contains the field stars and Class III sources 
(weak-line T Tauri stars). The region {}``T'' contains the CTTS \citep{lad92}, Class II 
objects with large NIR excesses, and reddened early type ZAMS (Zero Age Main Sequence) 
stars with \textit{K}-band excess emission. The {}``P'' zone
sources are most likely Class I objects (protostar-like objects) with
circumstellar envelopes. 
The NIR CC diagram is likely to be contaminated by the foreground and 
background field stars. To assess this, we also plotted the CC diagram for the 
control field (Figure \ref{fig_NIR_CCD_control}). On comparison, we find that almost all our control 
field sources are confined to the left of the middle reddening vector, and a few
which are to its right are below the CTT locus. Thus, the {}``T'' and {}``P''
regions of the object field are mostly uncontaminated by the field stars, and hence the sources 
in these regions can be assumed to be YSOs. 

As there is considerable nebulosity in the region, it must be kept in mind that there could
be dust knots emitting at near and mid infrared
wavelengths. These could be mistakenly identified as Class I/II sources
in the CC diagram. Hence, this CC diagram has its limitations when
it comes to the identification of the YSOs, and the information from
it is only complementary to other techniques like spectroscopy, to
understand the nature of objects. 
The spatial distribution of the YSOs identified was used to 
pin down the star formation process going on in this region (see Section \ref{subsection_Spatial_distribution}).

\subsection{NIR color-magnitude diagram}

\begin{figure}
\centering
\subfigure
{
\includegraphics[width=3.1in]{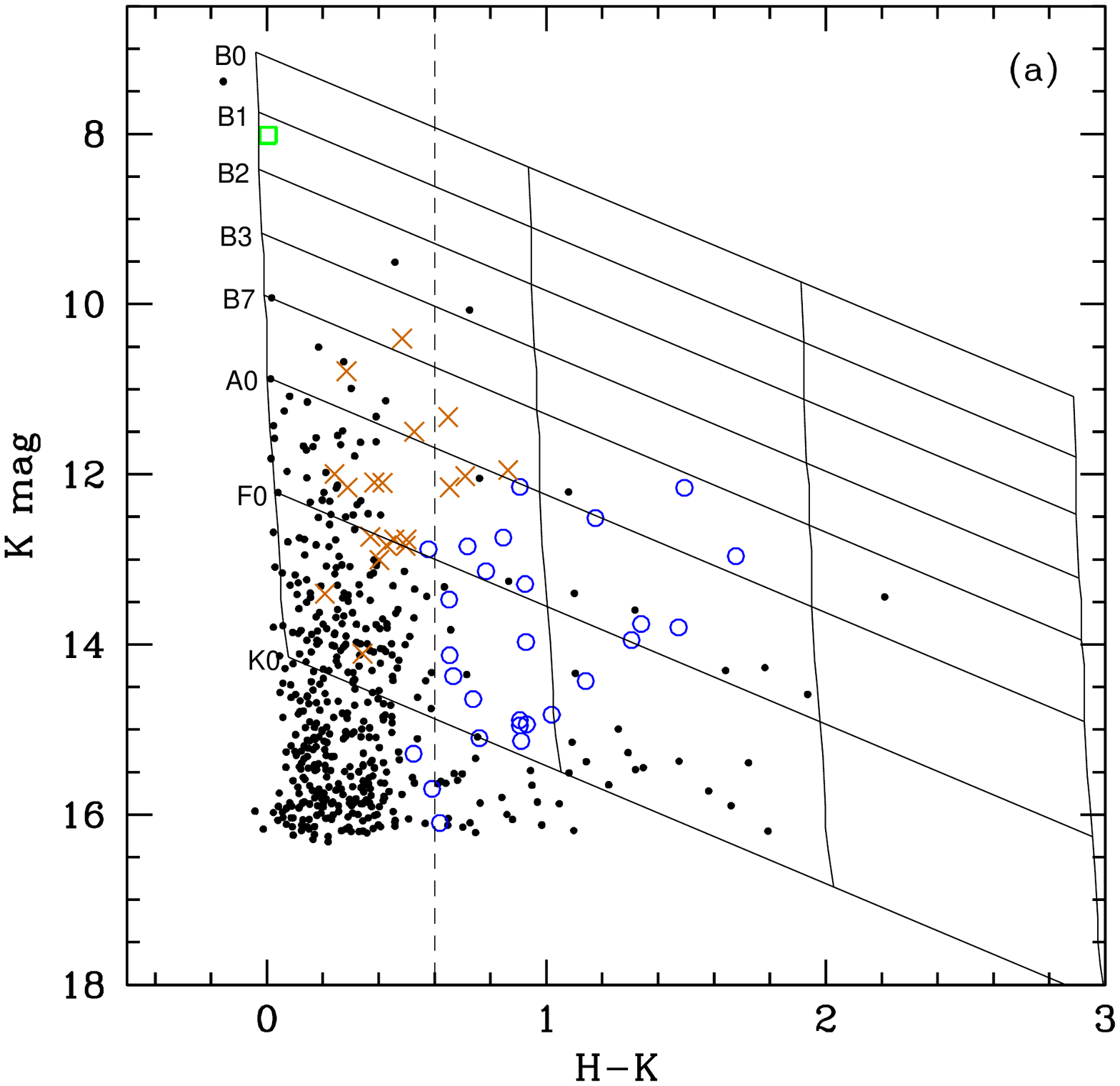}
\label{fig_NIR_CMD_object}
}
\subfigure
{
\includegraphics[width=3.1in]{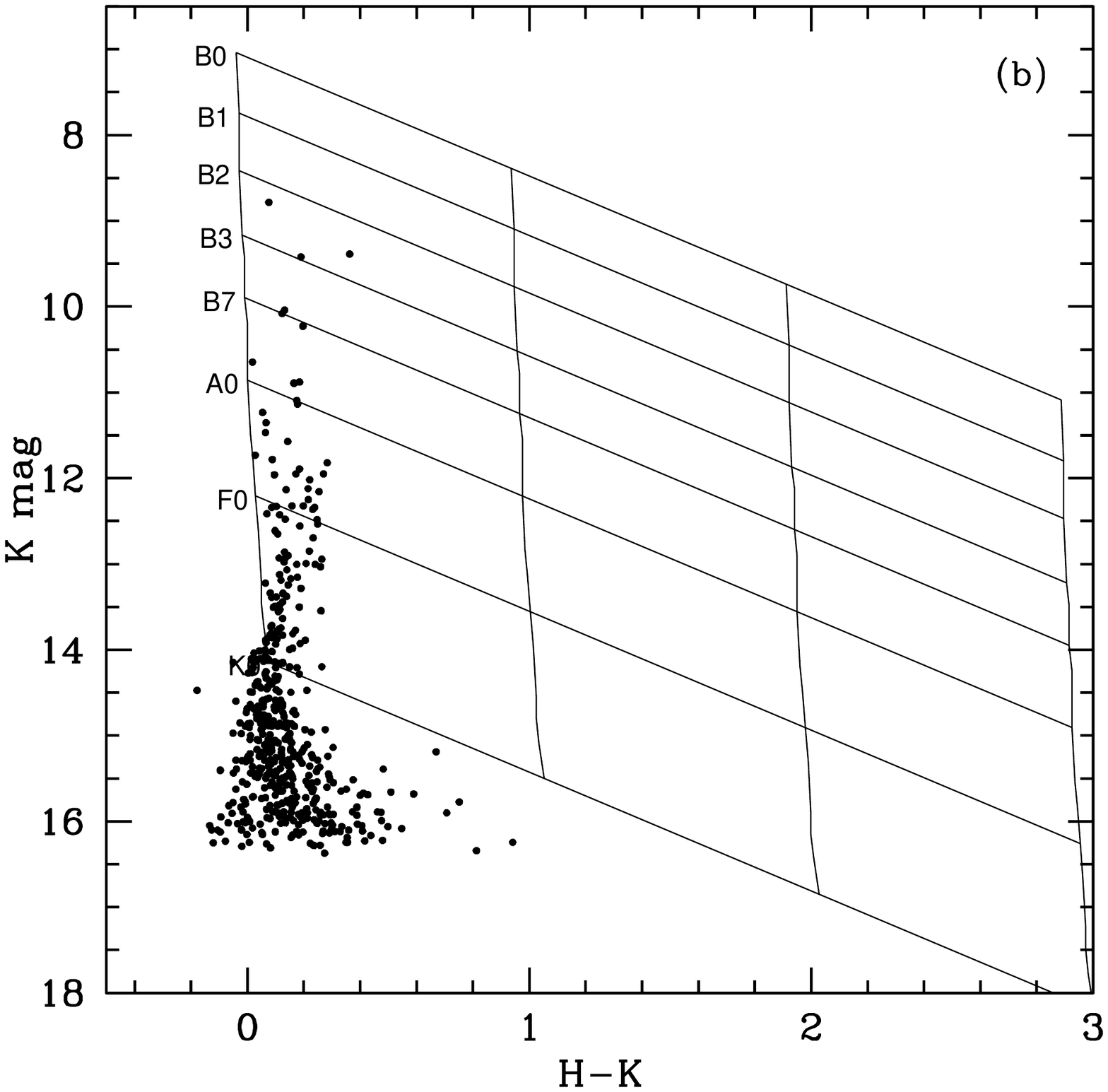}
\label{fig_NIR_CMD_control}
}
\caption{(a) NIR CM diagram for our object field. The nearly vertical solid lines are the loci of ZAMS stars reddened
by $A_{V}$=0, 15, 30, \& 45 mag. The parallel, slanting lines denote the reddening vectors.
Blue circles denote the Class II-like objects, orange crosses the H$\alpha$ emission line stars,
and the green square the central ionizing star. The $H-K$ cut-off at 0.6 has been marked with a dashed vertical line.
(b) CM diagram for the control field.}
\label{fig_NIR_CMD}
\end{figure}

The NIR CM diagrams, $K/(H-K)$, of the region and the control 
field are shown in Figure \ref{fig_NIR_CMD}.
The nearly vertical solid lines are the loci of 
ZAMS reddened by $A_{V}=0,\,15,\,30$ and $45$ mag. The slanting
lines are the reddening vectors for the corresponding spectral types.
A comparison of the object field and the control field CM diagrams  
shows an apparent separation between the stellar sources at around 
$H-K\sim$ 0.6. Our control field shows almost all sources having 
$H-K<0.6$, thereby indicating that sources with $H-K<0.6$ are most likely the 
foreground stars.
The sources lying to the right of the $H-K=0.6$ cut-off line 
(shown with a dashed vertical line in Figure \ref{fig_NIR_CMD_object}) have an 
excess emission in the \textit{K}-band and can thus reasonably 
be assumed to be young protostars. 
A CC diagram \citep[$K_{\mbox{s}}-3.4$/$3.4-4.6$,][]{koe12} for the 
red sources which had WISE counterparts (not shown here) showed their location in 
the diagram to be consistent with the location of Class I sources. 
The central ionizing star has been marked with a green square on this diagram. 
The NIR CM diagram indicates a B1 spectral type for the 
ionizing source, which is within one sub-class of that estimated from the optical spectroscopy
and radio continuum observations.
However, a spectral type derived from just two data points ($H$ and $K$) is not very reliable
because of uncertainty in distance determination. 
It just serves to give an idea of the spectral type, the spectrum analysis being canonical in this respect. 

We estimated the extragalactic contamination (extincted or faint
background galaxies) in our sample of NIR excess sources. 
We assume that the extragalactic sources can be seen through maximum 
extinction, 
which for our case is A$_{V} \sim$ 25 mag (A$_{K} \sim$ 2.25 mag) 
(see Section \ref{subsection_Extinction}).  
Accounting for the interstellar extinction and converting the \textit{K}$_{\rm S}$ magnitude 
to AB system (since \textit{K}$_{\rm S}$ counts are tabulated in AB 
photometric system in \citet{kee10}), the total number of galaxy counts in our NIR
region, i.e. $\sim$ 56.25 arcmin$^2$, turns out to be 6 $\pm$ 2 down to our 
faintest detected magnitude of $\sim$ 17 in the \textit{K}$_{\rm S}$ band. 

A list of all the YSOs, with their 
respective magnitudes in optical and IR, has been given in Table \ref{table_YSOs}. 
97 YSOs were analyzed in total, with 93 of them in our NIR FoV of $\sim$ 7.5$^{'} \times$ 7.5$^{'}$, 
and 4 H$\alpha$ emission line stars outside this FoV (but considered for analysis nevertheless). 
Out of the 93 YSOs, 15 are H$\alpha$ emission line stars detected using grism slitless spectroscopy with 
a \textit{V} limiting magnitude of 20.07, 76 are NIR YSOs (26 Class II-type sources $+$ 50 red 
sources with $H-K>0.6$) with a limiting $K_{\mbox{s}}$ magnitude of $\sim$ 17, and 2 are LDN1657A-2 and 
LDN1657A-3 sources from \citet{lin10} (only LDN1657A-3 was detected in NIR). 
Class II-like and $H-K>0.6$ sources which were also detected as H$\alpha$ emission line sources 
were classified as H$\alpha$ emission line sources. The $H-K>0.6$ sources which overlapped with 
Class II-like sources were classified as Class II-like sources.


\begin{deluxetable*}{c c c c c c c c c c c c c c c}
\tablecolumns{15}
\tablewidth{0pt}
\tablenum{2}
\tablecaption{Optical \& infrared magnitudes for the YSOs}
\startdata
\cline{1-15}
\multicolumn{15}{c}{Table 2 has been given at the end of the paper} \\
\enddata
\end{deluxetable*}

\subsection{Extinction in the region}
\label{subsection_Extinction}

The average extinction in the region was calculated using the CC diagram 
as well as the extinction map of the region.
 
To calculate the average reddening using the CC diagram,
we used the sources in the {}``F'' and {}``T'' zones of our diagram. The
sources in the {}``T'' region were dereddened to the CTT locus, while
the sources in the {}``F'' region were dereddened to the
MS locus which was approximated by a straight line asymptote. 
This gives the color-excess for each source. Subsequently, the reddening
laws of \citet{coh81} were followed to 
calculate the $A_{V}$ for each source. Finally,
we fit the set of $A_{V}$ values (ranging from $\sim$ 0.02 to 16 mag) obtained for each of the sources,
with a normal distribution. The mean of this distribution was
found to be 2.70 mag. 

The average visual 
extinction for the region was also found by making an extinction map
\citep[{}``NICE'' method,][]{lad94,kai07}. 
To do this, we first obtained the intrinsic $H-K$ color
from the control field. 
Thereafter, the color-excess, E($H-K$), and then $A_{V}$
for each of the sources in our catalog was calculated. The color-excess
is related to visual extinction by : E($H-K$) = 0.065$\times$ $A_{V}$,
which can be calculated from the reddening laws. 
$A_{V}$ values ranged from 0$-$25 mag, as we took all sources with 
at least \textit{H} and \textit{K}
magnitudes, and thereby even redder ones (undetected in \textit{J}), here. 
Subsequently, the extinction map was made, where the extinction at each pixel was
calculated by a weighted Gaussian mean around it. For our case, we 
took a Gaussian with an FWHM of 1$^{'}$. Finally, the
mean pixel value for our map was calculated using the task {}``IMSTAT''
in IRAF, which gave us an average visual extinction of 2.69 
mag.

Hence, we find similar $A_{V}$ values from both the methods. 
We used 2.70 mag as our value for $A_{V}$ in the optical CM diagram to estimate the mass and 
age of the YSOs (see Section \ref{subsection_Optical_CMD}). It gives the mean extinction of the field. 
However, owing to the nebulosity in the central region and the 
presence of dark cloud to the west (see Section \ref{subsection_Spatial_distribution}), 
individual extinction for the sources will vary depending on their location.

\subsection{Optical color-magnitude diagram}
\label{subsection_Optical_CMD}

\begin{figure}
\includegraphics[scale=0.4]{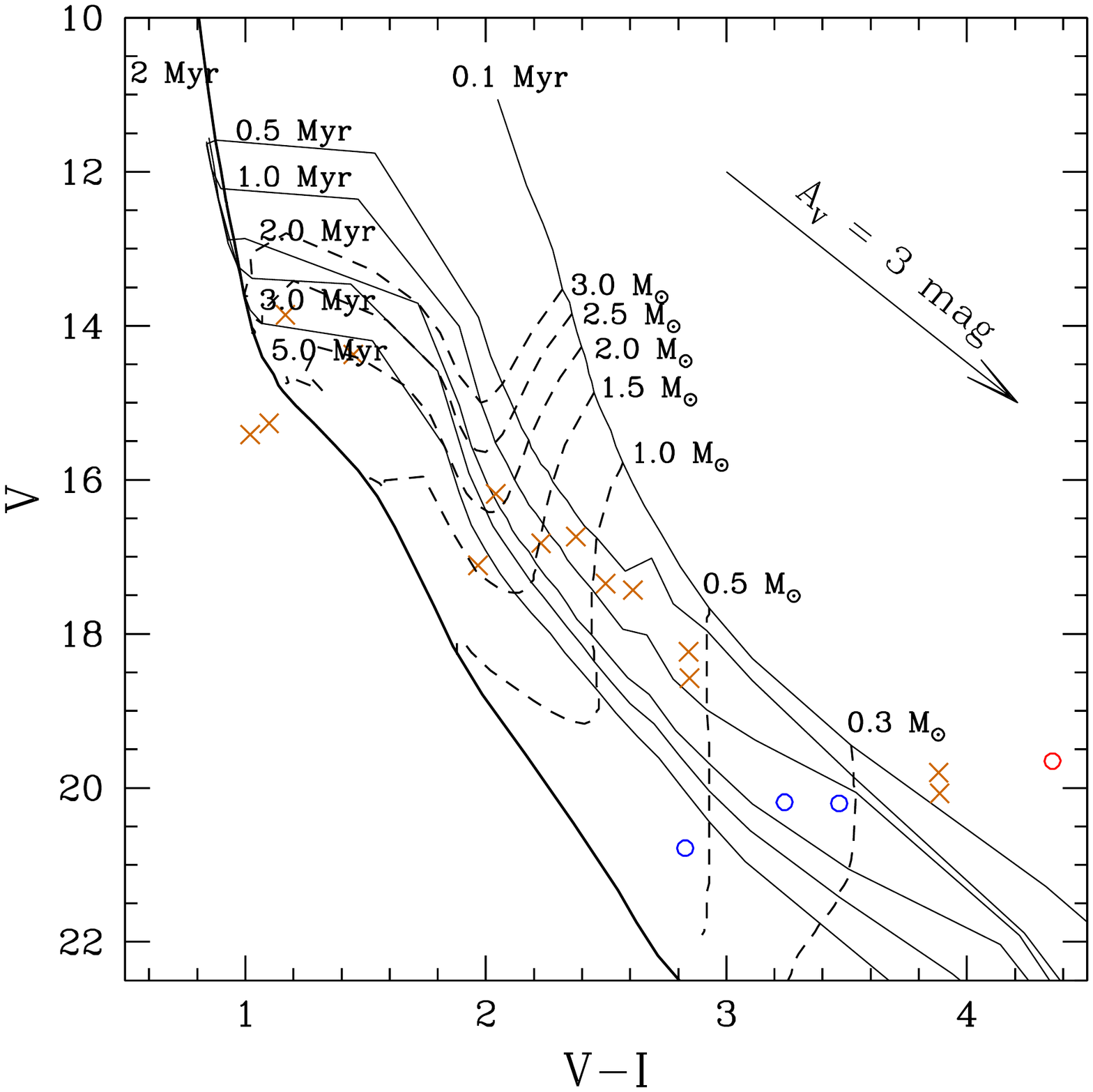}
\caption{Optical CM diagram showing H$\alpha$ emission line stars (orange crosses), Class II-type objects (blue circles),
and $H-K>0.6$ source (red circle). The thick
solid line is the 2 Myr isochrone by \citet{gir02}, while the thinner solid lines are PMS isochrones
for different ages by \citet{sie00}. Dashed lines are evolutionary tracks for different masses by \citet{sie00}.
The isochrones have been corrected for distance (1.1 kpc) and reddening ($A_{V}$ = 2.70 mag). The arrow shows
the reddening vector for $A_{V}$ = 3 mag.}
\label{fig_Optical_CMD}
\end{figure}

Figure \ref{fig_Optical_CMD} shows the \textit{V/V-I} CM diagram for the YSOs 
in the region. Here, H$\alpha$ emission line sources have been shown in orange crosses, 
Class II-like sources in blue circles, and sources with $H-K>0.6$ have been represented by red circles. 
Most of the sources detected at optical wavelengths are H$\alpha$ emission line sources, 
with only three Class II-like sources and one source with $H-K>0.6$. This is expected as H$\alpha$ emission line
sources are the most evolved YSOs in the field. 
In Figure \ref{fig_Optical_CMD}, the isochrone for 2 Myr for solar metallicity by \citet{gir02}, 
and pre-main sequence (PMS) isochrones for 0.1 to 5 Myr from \citet{sie00} 
have been overlaid. The isochrones were corrected for distance (1.1 kpc)
and reddening ($A_{V}$ = 2.70 mag; see Section \ref{subsection_Extinction}). The figure also shows the 
evolutionary tracks for masses ranging from 0.3 to 3 M$_{\odot}$ from \citet{sie00}. 

We can see that most of the YSOs lie in the age range of 0.5 to 2 Myr, with 
the mean age $\sim$ 1 Myr. The mass range mostly varies from 0.3 to $\sim$ 2 M$_{\odot}$, 
with a few outliers mainly being the H$\alpha$ emission line sources.  
The PMS star to the right of 0.1 Myr isochrone is likely to be 
a highly extincted source. It should be noted that variable extinction 
along the line of sight and variability could be the reason behind the age 
spread in our figure. Similar age spreads have been noted in other 
star-forming regions \citep{jos08,sha07}.

\subsection{Mass spectrum of the stellar sources}

\begin{figure}
\includegraphics[scale=0.4]{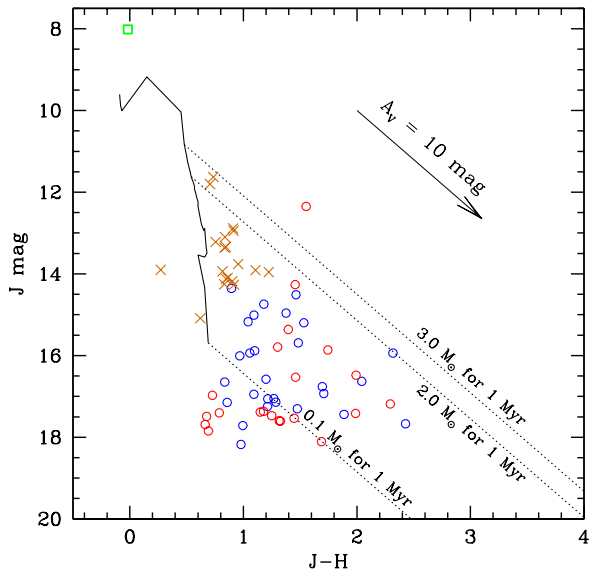}
\caption{\textit{J/J-H} CM diagram for the YSOs.
Blue circles denote the Class II-type sources, while red circles denote those with $H-K>0.6$. The
H$\alpha$ emission line stars have been represented by orange crosses. The central ionizing star
has been marked by a green square. The PMS isochrone for 1 Myr (solid line)
from \citet{sie00} has been shown. The isochrone was corrected for a distance of 1.1 kpc.
The slanted lines show the reddening vectors for different masses.}
\label{fig_Mass_spectrum}
\end{figure}

The excess \textit{K}-band emission 
can make the YSOs appear luminous and hence the $K/H-K$
CM diagram is not appropriate for mass estimation. We therefore
choose the $J/J-H$ CM diagram to minimize the effect of this excess
emission which would lead to an over estimation of mass. Figure \ref{fig_Mass_spectrum}
shows the required CM diagram, with the Class II-like sources detected from the
CC diagram (in blue circles), sources with $H-K>0.6$ (in red circles), 
and sources with H$\alpha$ emission line (in orange crosses). The ionizing 
star has also been shown with a green square. 
The larger statistics here as opposed to the optical CM diagram is 
due to the young nature of the sources, which emit at longer  
wavelengths. From the optical CM diagram, we found that the average 
age of the YSOs was about 1 Myr. Also, the dynamical 
age of this H~{\sc ii} region from radio results is 1.07 $\pm$ 0.31 Myr 
(see Section \ref{subsection_Physical_properties}). 
Hence, we use the isochrone for 1 Myr (black solid line) from \citet{sie00} 
for our analysis. 
The slanting dotted lines show the reddening vectors for different masses.

We can see from the 1 Myr reddening vectors, that most of the
H$\alpha$ emission line objects and Class II-like sources lie between 0.1 to 2 M$_{\odot}$, though a few
outliers seem to be much more massive, of the order of $\sim$ 3 M$_{\odot}$.
Most $H-K>0.6$ sources have masses less than 2 M$_{\odot}$.
It is noteworthy that our mass range is consistent with the mass range 
derived from the optical CM diagram (Section \ref{subsection_Optical_CMD}), 
with the H$\alpha$ emission line stars being towards the massive end.  
As can clearly be seen, the YSOs with the assumed age, 
have widely varying colors. This is probably an indication of variable
extinction, weak contribution of excess emission in \textit{J} and \textit{H}
bands, and sources being in different evolutionary stages. It should
be noted that mass estimations from infrared CM diagrams are prone to
systematic errors \citep{hil08} due to possible use of different PMS evolutionary models, and
reliance on uncertain age and distance. Another error source could be the 
presence of binaries, which will brighten a star - thus changing the mass and age estimates.

\subsection{Spectral energy distribution}
\label{section_SED}

So far, we have identified the YSOs with the help of NIR CC and CM
diagrams, and the H$\alpha$ emission line stars. 
To get an idea about the physical parameters of these sources, 
we modelled the SEDs using the grid of
models and fitting tools of \citet{rob06,rob07}.

\begin{figure*}
\centering
\vspace*{-6cm}
\subfigure
{
\includegraphics[scale=0.4]{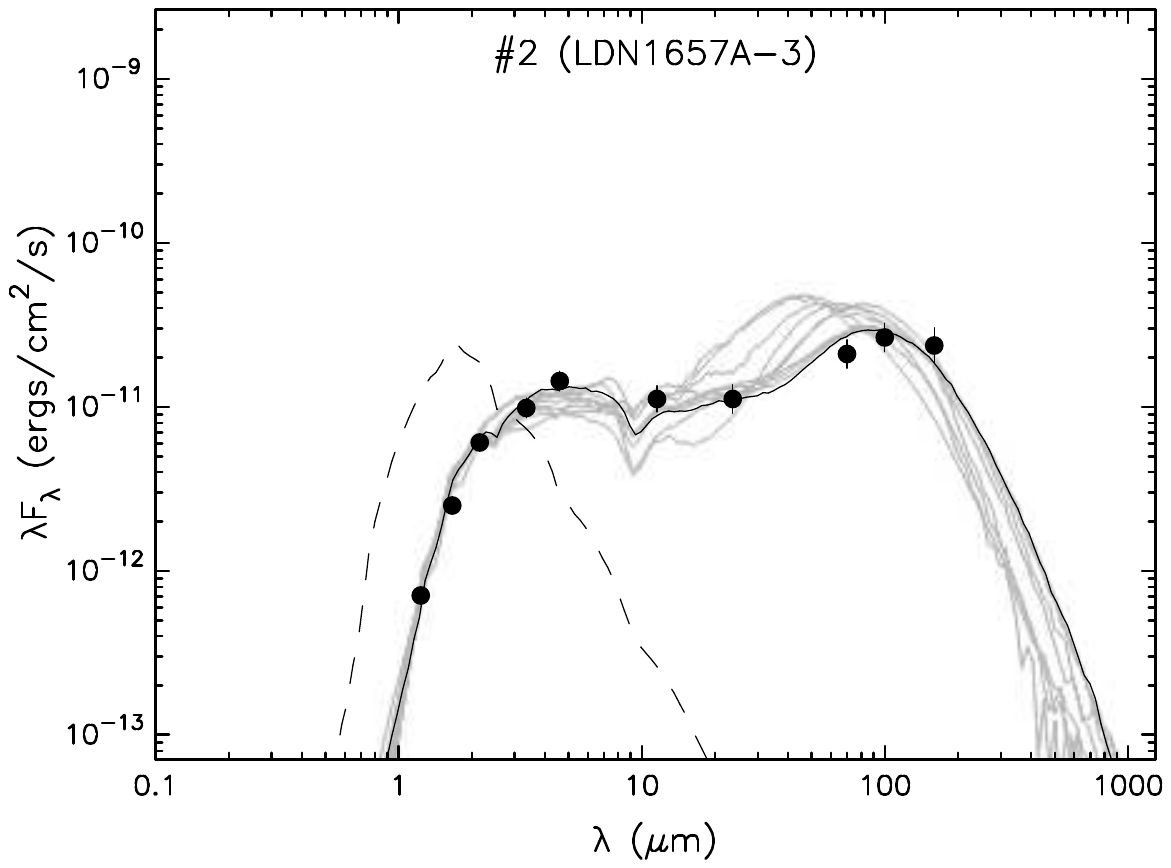}
\label{fig_linz_SED}
}
\subfigure
{
\includegraphics[scale=0.4]{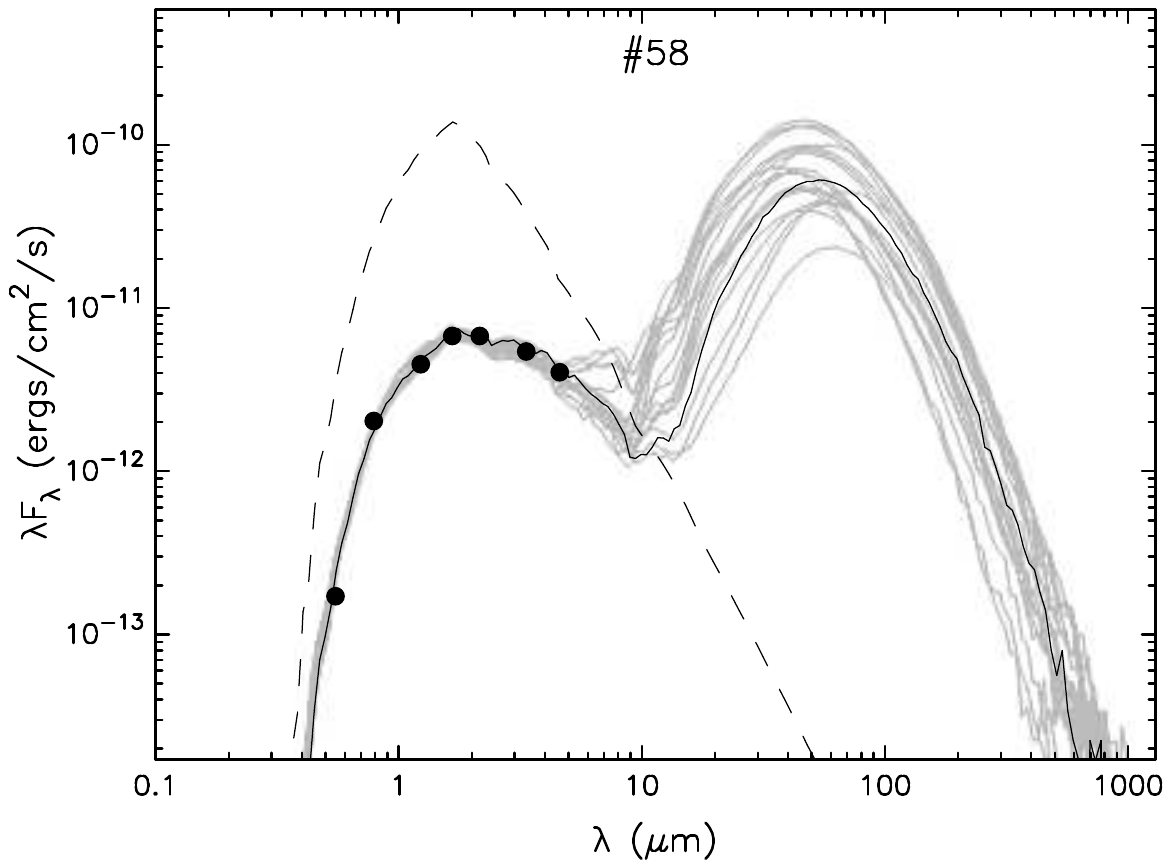}
\label{fig_CTT_SED}
}
\caption{SED fitting using the model of \citet{rob07} for (\textit{left}) LDN1657A-3, from \citet{lin10}, and
(\textit{right}) a typical Class II-type source.
The ID no. corresponds to the IDs in the YSO table (Table \ref{table_YSOs}). The black dots are our the data points.
The solid black line shows the best fitted model, while the grey lines
show the consecutive good fits for $\chi^{2} - \chi^{2}_{min} ~ (per~data~point) < 3$. 
The dashed line shows the fit for the photosphere of the central source for the
best fit model. The photosphere fit is assuming that there is no circumstellar dust, however, interstellar reddening has
been accounted for.}
\label{fig_SED}
\end{figure*}

The basic model assumes three components of any 
source : a PMS central star, a surrounding flared
accretion disk, and a rotationally flattened envelope with cavities carved
out by a bipolar outflow. 
Different models are computed for a suitably large parameter space - 
containing manifold combinations of the above mentioned components - using a 
Monte Carlo based radiation transfer code \citep{whi03a,whi03b}. 
SED fitting for different sources is not unique and the number of models
for each source can only be constrained by having more data points
from multiwavelength observations. 
To reduce error due to the uncertainties which result from very few data 
points, we modelled only those sources for which we had WISE data and at least 
five data points. 
The input parameters were distance range (taken to be 0.9$-$1.3 pc), 
and A$_{V}$ (taken as 1.1$-$25 mag). 
The lower limit was chosen to be 1.1 mag (from the calculation of \citet{bon09}
for \object{BDSB 96}) as the YSOs should have an extinction 
which is more than the visual absorption for the region. The upper limit 
was chosen as 25 mag because the highest individual A$_{V}$ values from 
color-excess calculations using NICE method (Section \ref{subsection_Extinction}) were around this value.    
The SED fitting tool of \citet{rob07} gives a set of 
well constrained values for each parameter, rather than the values 
for only the best fit for SED. Following a similar method as 
\citet{rob07}, we considered only those fits to constrain the physical 
parameters for which we had 

\begin{equation}
\label{equation_SED}
 \chi^{2} - \chi_{min}^{2} < 3 \, \mbox{(per data point)},
\end{equation} 

where $\chi^{2}$ is measure of goodness of fit for each model (see Figure \ref{fig_SED}). The final 
values were subsequently obtained using the models which satisfied 
Equation \ref{equation_SED} and finding a weighted mean value for each parameter, with the 
weight being the inverse square of $\chi^{2}$ for each model respectively. 
For some of the sources, the errors for a few parameters are large because 
we are dealing with a large parameter space, while we have only a few data
points to fit the models. 
The results from our SED modelling are given in Table \ref{table_SED}. The physical 
parameters listed are stellar mass ($M_{\odot}$), stellar age ($t_{*}$),
temperature ($T$), disk accretion rate ($\dot{M}_{disk}$), 
foreground visual extinction ($A_{V}$), and the $\chi^{2}_{min}$ per data point of the best fit. 

\begin{deluxetable*}{c c c c c c c c c}
\tablecolumns{9}
\tablewidth{0pt}
\tablenum{3}
\renewcommand{\arraystretch}{1}
\tabletypesize{\tiny}
\tablecaption{SED results for the YSOs \label{table_SED}}
\tablehead{
\colhead{ID\tablenotemark{a}} &   \colhead{RA} &   \colhead{Dec.}       &    \colhead{Mass}      &    \colhead{$\log t_{*}$}     &    \colhead{$\log T$}        
  & \colhead{$\log\dot{M}_{disk}$} & \colhead{$A_{V}$}   & \colhead{$\chi_{min}^{2}$} \\
\colhead{}   &   \colhead{(J2000)}      &   \colhead{(J2000)}    &  \colhead{(M$_{\odot}$)}      &    \colhead{(yr)}   
&    \colhead{(K)}     &   \colhead{(M$_{\odot}$ yr$^{-1}$)}  & \colhead{(mag)}  &   \colhead{}}
\startdata
\multicolumn{9}{c}{Sources from Linz et al. (2010)} \\
\cline{1-9}        
 1  &    07:04:58.030  &    -12:16:50.80  &       0.21 $\pm$  0.39  &       3.37 $\pm$  0.46  &       3.46 $\pm$  0.07  &      -6.82 $\pm$  0.55  &	 16.23 $\pm$  7.82  &	4.36	\\
 2  &    07:05:00.665  &    -12:16:45.13  &       1.27 $\pm$  1.49  &       5.32 $\pm$  0.39  &       3.61 $\pm$  0.10  &      -7.71 $\pm$  1.85  &	 10.56 $\pm$  4.20  &	1.89	\\
\cline{1-9}
\multicolumn{9}{c}{Sources with $H-K>0.6$} \\
\cline{1-9}
19  &    07:05:00.919  &    -12:15:42.37  &       2.44 $\pm$  1.00  &       5.43 $\pm$  0.44  &       3.67 $\pm$  0.05  &      -9.15 $\pm$  1.97  &	  7.27 $\pm$  1.14  &	1.40	\\
45  &    07:05:10.395  &    -12:19:25.70  &       2.91 $\pm$  1.38  &       4.55 $\pm$  1.27  &       3.81 $\pm$  0.27  &      -6.35 $\pm$  1.45  &  1.89 $\pm$  0.78  &   8.28   \\
49  &    07:05:13.678  &    -12:19:51.96  &       1.14 $\pm$  0.72  &       5.05 $\pm$  0.38  &       3.61 $\pm$  0.03  &      -8.18 $\pm$  1.11  &  2.90 $\pm$  1.83  &   1.20	\\
\cline{1-9}
\multicolumn{9}{c}{Class II-type sources} \\
\cline{1-9}
53  &    07:04:56.673  &    -12:16:12.77  &       0.97 $\pm$  0.82  &       5.21 $\pm$  0.37  &       3.60 $\pm$  0.03  &      -8.74 $\pm$  1.20  &	  2.22 $\pm$  1.16  &	3.23	\\
54  &    07:04:57.964  &    -12:17:13.85  &       0.19 $\pm$  0.22  &       5.90 $\pm$  0.48  &       3.49 $\pm$  0.03  &     -10.05 $\pm$  1.46  &	  3.94 $\pm$  0.88  &	1.45	\\
58  &    07:05:01.230  &    -12:19:03.69  &       1.28 $\pm$  0.90  &       5.15 $\pm$  0.51  &       3.62 $\pm$  0.04  &      -9.01 $\pm$  1.41  &	  2.98 $\pm$  1.32  &	3.34	\\
64  &    07:05:08.269  &    -12:19:15.83  &       3.62 $\pm$  2.24  &       5.37 $\pm$  1.20  &       3.87 $\pm$  0.30  &      -8.82 $\pm$  2.73  &	 15.93 $\pm$  5.87  &	0.34	\\
66  &    07:05:09.197  &    -12:19:00.00  &       0.86 $\pm$  0.74  &       4.63 $\pm$  0.72  &       3.59 $\pm$  0.09  &      -7.88 $\pm$  1.20  &	  4.72 $\pm$  2.26  &	1.64	\\
69  &    07:05:11.424  &    -12:19:24.36  &       0.86 $\pm$  0.71  &       4.72 $\pm$  0.65  &       3.59 $\pm$  0.08  &      -7.98 $\pm$  1.15  &	  4.45 $\pm$  1.95  &	1.50	\\
70  &    07:05:11.541  &    -12:18:11.39  &       1.32 $\pm$  1.35  &       4.43 $\pm$  0.79  &       3.60 $\pm$  0.06  &      -7.93 $\pm$  1.12  &  3.07 $\pm$  1.37  &   2.21    \\
72  &    07:05:11.774  &    -12:16:24.01  &       1.00 $\pm$  0.80  &       5.24 $\pm$  0.58  &       3.64 $\pm$  0.13  &      -8.56 $\pm$  1.45  &	  2.71 $\pm$  1.52  &	2.13	\\
73  &    07:05:14.495  &    -12:20:47.57  &       0.53 $\pm$  0.79  &       4.93 $\pm$  0.89  &       3.52 $\pm$  0.14  &      -8.76 $\pm$  1.21  &	  2.45 $\pm$  1.19  &	0.03	\\
75  &    07:05:15.324  &    -12:20:02.05  &       0.84 $\pm$  1.09  &       4.42 $\pm$  0.90  &       3.58 $\pm$  0.15  &      -7.45 $\pm$  1.23  &  2.60 $\pm$  1.92  &   2.15   \\
76  &    07:05:17.419  &    -12:18:58.69  &       0.81 $\pm$  1.19  &       4.85 $\pm$  1.15  &       3.56 $\pm$  0.22  &      -8.20 $\pm$  1.42  &	  2.21 $\pm$  1.01  &	0.04	\\
\cline{1-9}
\multicolumn{9}{c}{H$\alpha$ emission line sources} \\
\cline{1-9}
79  &    07:05:06.330  &    -12:15:36.38  &       2.27 $\pm$  0.67  &       6.63 $\pm$  0.36  &       3.85 $\pm$  0.15  &      -9.95 $\pm$  1.95  &	  5.11 $\pm$  1.05  &	1.91	\\
80  &    07:05:09.652  &    -12:19:56.32  &       0.89 $\pm$  1.18  &       4.60 $\pm$  0.82  &       3.58 $\pm$  0.06  &      -7.90 $\pm$  0.98  &	  3.24 $\pm$  1.59  &	1.15	\\
82  &    07:05:12.292  &    -12:18:38.27  &       0.94 $\pm$  0.92  &       5.68 $\pm$  0.30  &       3.61 $\pm$  0.04  &      -8.76 $\pm$  1.15  &	  1.46 $\pm$  0.52  &	1.52	\\
86  &    07:05:17.904  &    -12:15:40.26  &       4.60 $\pm$  1.03  &       5.61 $\pm$  0.07  &       3.78 $\pm$  0.09  &      -7.48 $\pm$  1.00  &	  2.01 $\pm$  0.60  &	2.62	\\
87  &    07:05:18.494  &    -12:15:09.22  &       2.30 $\pm$  0.72  &       6.04 $\pm$  0.58  &       3.79 $\pm$  0.17  &      -9.08 $\pm$  1.49  &	  2.31 $\pm$  1.05  &	0.54	\\
88  &    07:05:19.389  &    -12:20:28.71  &       2.51 $\pm$  0.81  &       6.55 $\pm$  0.31  &       3.82 $\pm$  0.16  &      -9.41 $\pm$  2.01  &	  2.23 $\pm$  0.92  &	0.24	\\
89  &    07:05:19.592  &    -12:17:31.48  &       2.33 $\pm$  0.86  &       5.58 $\pm$  0.42  &       3.67 $\pm$  0.05  &      -9.61 $\pm$  1.32  &  2.12 $\pm$  0.58  &   2.68   \\
90  &    07:05:20.068  &    -12:19:12.59  &       3.93 $\pm$  0.15  &       6.63 $\pm$  0.13  &       4.15 $\pm$  0.01  &     -12.90 $\pm$  0.60  &  3.83 $\pm$  0.21  &   1.27   \\
91  &    07:05:20.728  &    -12:21:42.69  &       2.64 $\pm$  0.78  &       5.57 $\pm$  0.31  &       3.68 $\pm$  0.05  &      -9.15 $\pm$  1.64  &	  1.91 $\pm$  0.66  &  1.04	\\
92  &    07:05:21.912  &    -12:19:07.19  &       0.26 $\pm$  0.18  &       5.83 $\pm$  0.29  &       3.51 $\pm$  0.02  &     -10.70 $\pm$  1.64  &	  1.34 $\pm$  0.34  &	2.61	\\
93  &    07:05:22.359  &    -12:17:18.63  &       1.07 $\pm$  0.88  &       6.14 $\pm$  0.45  &       3.62 $\pm$  0.06  &      -9.20 $\pm$  1.18  &	  2.06 $\pm$  1.00  &	1.25	\\
\cline{1-9}
\multicolumn{9}{c}{H$\alpha$ emission line sources outside NIR FoV} \\
\cline{1-9}
96  &    07:05:23.669  &    -12:22:36.08  &       1.69 $\pm$  0.64  &       6.22 $\pm$  0.38  &       3.71 $\pm$  0.13  &      -9.96 $\pm$  1.58  &	  2.46 $\pm$  1.11  &	1.64	\\  
\enddata
\tablenotetext{a}{The ID no. refers to the ID in Table \ref{table_YSOs}.}
\end{deluxetable*}

Two of the PMS sources detected by \citet{lin10} - LDN1657A-2, \& 
LDN1657A-3 - had WISE counterparts. LDN1657A-3 was also detected at 
NIR wavelengths. Using the flux values given for these sources at longer  
wavelengths from \citet{lin10}, we could further constrain the 
physical parameters of these sources. The source with ID {}``1'' in our SED table 
(Table \ref{table_SED}) gives the values for LDN1657A-2, while {}``2'' gives those for LDN1657A-3. 
Figure \ref{fig_SED} shows examples of SEDs with the resulting models 
for LDN1657A-3 and a typical Class-II like source. 
 
It must be kept in mind that the SED results are only representative of the actual values
and should not be taken too literally. Out of a database of 200000 YSO models (each for a different 
set of parameter values), the model simply finds the closest matches for a particular source entered
by the user. Any grid of models will have inherent limitations, which could give rise to pseudo-trends 
in the results. The models are for isolated objects and thus could be misleading when we have unresolved 
multiple point sources. Moreover, the model assumes same physics for stellar masses from 0.1 to 50 M$_{\odot}$. 
\citet{rob08} provides a succinct summary of the caveats inherent in this modelling.  
Nevertheless, we can still use the results to glean general information about the mass and age ranges 
of our sources. 
As can be seen from Table \ref{table_SED}, most of the sources have 
stellar ages in the range 0.05 to 1 Myr and masses in 0.2 to 2.5 M$_{\odot}$ range.
The outliers are expectedly the more evolved H$\alpha$ emission line stars, whose ages vary 
upto 5 Myr, and stellar masses are $\sim$ 2.5$-$5 M$_{\odot}$. We see a few 
very low mass (3 sources with masses $<$ 0.5 M$_{\odot}$) and extremely young sources 
(9 sources with ages $< 10^{5}$ yr, out of which 6 are CTTS) as well. 
The disk accretion rate for most of the H$\alpha$ emission line sources is in the range of $\sim$ 
10$^{-9}$-10$^{-11}$ M$_{\odot}$ yr$^{-1}$, which is less than that of CTTS $\sim$ 
10$^{-7}$-10$^{-9}$ M$_{\odot}$ yr$^{-1}$, though it would be prudent to treat these values 
with caution owing to a lack of FIR and submillimetre data points. 
A few of our sources, ID no.s 39, 55, 61, 81, 83, 94, 95, and 97 (Table \ref{table_YSOs}), 
were better fitted by stellar models of 
a heavily extincted central star, and are thus not included here.
Hence, we find that our age and mass estimations are mostly consistent with 
those derived from the optical CM diagram and the mass spectrum above.

\subsection{Spatial distribution of the YSOs}
\label{subsection_Spatial_distribution}

\begin{figure}
\includegraphics[scale=0.4,angle=90]{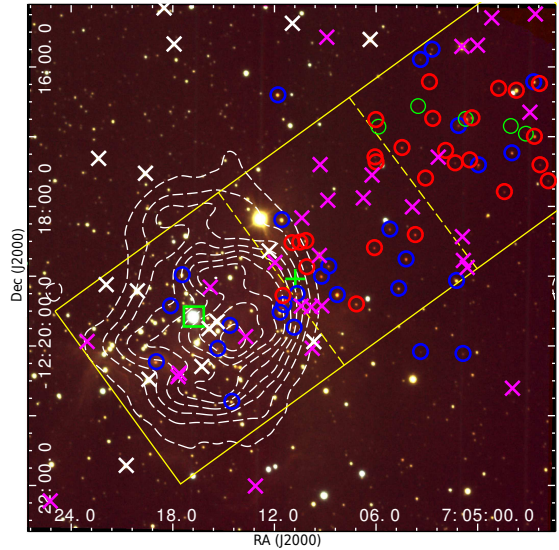}
\caption{Color composite showing the spatial distribution of the YSOs,
made using IRSF \textit{J} (blue), \textit{H} (green), and $K_{\rm S}$ (red) band images.
The H$\alpha$ emission line stars have been denoted by
white crosses, Class II-like sources by blue circles, sources with $1 \geq H-K>0.6$ with magenta crosses, and sources
with $H-K>1$ by red circles. The green square marks the position of the central ionizing star HD 53623.
UYSO1 \citep{for04} has been marked with a green plus sign, while the cool, compact sources from \citet{lin10}
by green circles. 1280 MHz GMRT contours ($\sim$ 25$^{''}$ resolution) showing the champagne
flow have been shown in dashed white lines. The three sub-regions denoted by
the yellow boxes show an increasing
fraction of redder sources in the sub-regions as we move from the
ionizing star towards the north-western direction. The dashed yellow lines are drawn at
a distance of 80$^{''}$ and 220$^{''}$ from the ionizing star, respectively.}
\label{fig_NIR_ColorComposite_with_sources}
\end{figure}

Figure \ref{fig_NIR_ColorComposite_with_sources} shows the spatial distribution of the YSOs from 
our IRSF and grism slitless spectroscopy data. Red circles 
denote the sources with $H-K>1$, sources with $1 \geq H-K>0.6$ 
have been shown with magenta crosses, blue circles 
denote the Class II-like objects detected from our CC diagram, while the white crosses 
show the location of the H$\alpha$ emission line stars from grism slitless
spectroscopy. The central ionizing source \object{HD 53623} is shown with a 
green square. \citet{for04} had found a young, intermediate mass protostar  
named UYSO1 at the north-west 
border of the H~{\sc ii} region, towards the cold dark cloud \object{LDN1657A}. The 
location of UYSO1, which was further resolved into two protostars 
\citep{for09}, has been denoted by a green plus sign. 
In addition, a recent Herschel PACS and SPIRE mapping of the region around UYSO1
identified five cool and compact FIR sources, of masses of the 
order of a few solar masses, hidden in the dark cloud \citep{lin10}. 
They have been marked by green circles.
Most of the YSOs identified by us lie in the molecular cloud region traced 
by $^{12}$CO molecular emission (see Section \ref{subsection_IR_and_submm_structures} \& Figure \ref{fig_CO3to2}),
along the concave arc facing the north-eastern corner. 

\begin{figure}
\includegraphics[scale=0.4]{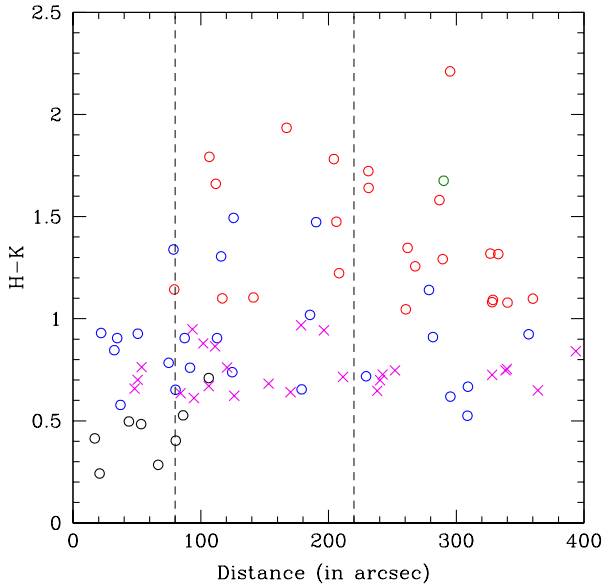}
\caption{The variation of $H-K$ color with distance from the star HD 53623.
The H$\alpha$ emission line stars have been denoted by
black circles, Class II-like sources by blue circles, sources with $1 \geq H-K>0.6$ with magenta crosses, and sources
with $H-K>1$ by red circles. Green circle denotes LDN1657A-3 source.}
\label{fig_Distance_and_color}
\end{figure}

We have divided the region into three sub-regions which show increasing infrared excess   
from the south-east towards the north-west direction. This indicates an evolutionary sequence,
thereby suggesting a possible decreasing age gradient towards that direction.
The youngest stars, i.e. those with $H-K>0.6$ mostly 
lie towards the north-west direction, followed by a concentration of Class II-like objects 
and then the H$\alpha$ emission line stars towards the ionizing region.
Even among the stars with $H-K>0.6$, the redder sources ($H-K>1$) were found to be lying 
towards the north-west, with a decreasing color sequence towards the 
ionizing source. 
Also, the 5 sources found by Herschel mapping are cooler than the massive 
object UYSO1 \citep{lin10}, thereby indicating that they are younger than 
UYSO1.
We tried to quantify the infrared excess ($H-K$) for increasing distances of the sources 
from the ionizing star HD 53623. Figure \ref{fig_Distance_and_color} shows a plot of $H-K$ color 
{\it versus} distance. The graph is 
divided into three regions as shown by three boxes in Figure 12, 
with the average $H-K$ color calculated for each of them. The average 
color was found to increase as we move away from the ionizing star, thereby 
indicating that the sources in general,
are younger at larger distances from the central region. 

This distribution of YSOs, with relatively larger NIR excess at larger radial distances, 
could be indicative of probable
triggered star formation, which seems to have propagated from the 
ionizing source towards the north-west.

\section{Star formation scenario}
\label{section_StarFormation_Scenario}

Star formation in a molecular cloud is more active during the first few million 
years of cloud time. Subsequent to that, the gas in molecular cloud dissipates and 
further star formation no longer takes place. However, these types of molecular cloud 
complexes can still have star formation at the interface of the H~{\sc ii} region 
created by OB stars and the molecular gas. 
In \object{Sh2-297} region,  
the spatial distribution of the YSOs and infrared excess calculations (Section \ref{subsection_Spatial_distribution})
suggest an evolutionary sequence of sources from the ionizing region towards
the cold, dark cloud \object{LDN1657A}, with a majority of YSOs distributed at the 
western and north-western border of \object{Sh2-297}. 

One of the conditions for the massive star \object{HD 53623} 
to have influenced the formation of YSOs in the region
is that the age of the H~{\sc ii} region and associated sources should be of the order of, 
or greater than, the age of the YSOs in the triggered region. 
However, it is difficult to estimate the age of YSOs as very few of them have optical counterparts.  
Nonetheless, we tried to estimate the age of the few sources which 
have optical counterparts in Section \ref{subsection_Optical_CMD}.
Age estimation using IR data could be misleading because IR emission arises not only from  
the photosphere, but also the circumstellar disk of the YSOs. 
We tried to constrain the age of embedded YSOs with the SED fitting models, 
though due to the lack of sufficient data points, the estimated ages are uncertain. 
However, they can still be used as quantitative indicators of stellar youth. 
Our analysis shows the mean age of YSOs to be $\sim$ 1 Myr, which is of the order of the 
dynamical age of the H~{\sc ii} region of 1.07 $\pm$ 0.31 Myr, 
considering the uncertainties in age estimation. 
Our SED results (Table \ref{table_SED}) also 
point out that barring 6 sources (all being H$\alpha$ emission line sources; 
possibly associated with the evolved H~{\sc ii} region), the other 22 sources have ages less than 1.07 Myr
(majority of them are located in the dark cloud at the western border of \object{Sh2-297}). 
The protostellar source detected by \citet{for09}, and the cool, compact 
sources detected by \citet{lin10} also show recent star formation occurring 
in the dark cloud. The dynamical age of the outflow for UYSO1, which was 
constrained by \citet{for09} as $<$ 10$^{4}$ yr, and its location at the 
immediate interface of \object{Sh2-297} supports our notion of triggered star formation 
by the expanding H~{\sc ii} region. 

To examine if the collect-and-collapse due to the ionizing star is possibly at work in the region,
we compare the observed properties with the analytical model by 
\citet[Section 5]{whi94} for star formation at the periphery of an expanding H~{\sc ii} region. 
The model gives the expressions for the time when the 
fragmentation starts (t$_{frag}$), and the radius at the 
time of fragmentation (r$_{frag}$).
The value of  $a_{s}$ (isothermal sound speed in the shocked layer), which varies 
from 0.2 to 0.6 km s$^{-1}$, was taken to be 0.2 km s$^{-1}$ for our calculations. 
Using the values of $S_{*}$ and $n_{o}$ calculated in previous section  
(Section \ref{subsection_Physical_properties}), we obtained $t_{frag}$ $\approx$ 1.27 Myr and 
$r_{frag}$ $\approx$ 1.70 pc. 
This would suggest that fragmentation would start at a radius of 1.70 pc, 
whereas the stellar sources found are within a 1.70 pc radius of the 
massive star \object{HD 53623}. 
However, the distance of the YSOs is the projected distance 
and the actual distance could be larger.
Taking the present mean age of the YSOs as 
$\sim$ 1 Myr and the fragmentation time as calculated, it would imply that
the age difference between the YSOs and \object{HD 53623} ($\sim$ 5 $\pm$ 3 Myr) is $>$ 2 Myr. 
Though this $t_{frag}$ calculation suggests that this H~{\sc ii} region could be  
old enough to lead to star formation by collect-and-collapse process, 
the morphology of the region makes this process unlikely.
Moreover, we have assumed the lowest value of isothermal sound speed for our 
calculations (0.2 km s$^{-1}$). A higher value of $a_{s}$ will only further increase 
the fragmentation time and radius and make a collect-and-collapse scenario unlikelier. 
Also, the YSOs are not exclusively associated with the 
PAH emission that we see around this region (Figure \ref{fig_CO3to2}). PAH regions represent
the compressed neutral matter between the ionization and shock fronts.  
The YSOs extend well beyond the PAH emission, as is evident from our MSX 
8.28 $\mu$m contours (which, in addition, contains continuum emission too). 
This is distinctly separate from the 
star forming regions of, for example, \citet{deh05} where young 
sources and condensations have been observed along the PAH ring. 
It should be noted that, in an ideal case, one would expect 
regularly-spaced-clumps/protostars of very similar ages distributed 
along the periphery of the H~{\sc ii} region, mostly along the material swept-up 
by the expanding ionization front. This is definitely not the case here. 
It therefore appears that the collect-and-collapse process due to the star HD 53623 is most likely 
not involved in the star formation in this region. 

Alternatively, it is possible that the star formation 
observed at the north-western border of Sh2-297 is probably due to 
some other processes. 
The formation of YSOs in the dark cloud area could partly be due to the influence of 
expanding H~{\sc ii} region Sh2-297 by RDI (radiation driven implosion) mechanism 
(most likely upto the edge and its vicinity)
\citep{ber89,lef94}, and partly due to supernova induced star formation as discussed by \citet{her77} 
(the distant YSOs possibly having formed by this process). 
The YSOs are distributed along the concave arc (see Section \ref{subsection_Spatial_distribution})
of the molecular cloud traced by $^{12}$CO, and seem to have an age sequence. 
It appears that the supernova might have swept the 
molecular material and led to its piling up along the arc. Therefore, as the material might have  
collected along the arc, beginning from the edge of Sh2-297 towards the north-west, 
star formation could have occured at different epochs. 
The plausible location of the supernova 
\citep[$\alpha_{1950}$ $\sim$ 07$^{h}$08$^{m}$, $\delta_{1950}$ $\sim$ -11.2$^{o}$,][]{rey78}
and the derived age of the supernova remnant - 0.5 Myr \citep{her77} - lend weight to this 
hypothesis. 

To better understand the star formation and the effect of the 
massive star on the \object{Sh2-297} region, we need much
deeper, wider and longer wavelength observations for better statistics (so as to constrain mass and
luminosity functions of the embedded YSOs), and to carry out a study of YSOs along the arc of molecular
material connecting \object{Sh2-297} to the larger region. It would also be helpful to carry out 
a radial velocity study and spectroscopic observations for a wider area to 
constrain the membership, age and spectral type of the YSOs. This would confirm their 
relative evolutionary status', identify possible clusters, and help put the relevant
triggered star formation process on a firm footing.

\section{Conclusions}
\label{section_Conclusions}

In this paper, we have carried out a multiwavelength study of the 
star-forming region \object{Sh2-297}. 
Our main conclusions are as follows :
\begin{enumerate}
\item
We identified the spectral class of the central ionizing star \object{HD 53623}, using 
optical spectroscopic data.
The prominence of He I lines in our spectra indicate a spectral type of 
B0V. Our optical spectroscopic results are consistent with the  
radio continuum data results.   
\item
Using the radio data for 610, 1280, \& 1420 MHz, the physical parameters of the 
region were derived. Taking the extent of the H~{\sc ii} region to be 1.6 pc (5$^{'}$), 
the emission measure was calculated to be 
9.15 $\pm$ 0.56 $\times$ 10$^{5}$ cm$^{-6}$ pc, and thereby the electron density 
as 756 $\pm$ 46 cm$^{-3}$. These low values of emission measure and the electron 
density confirm that this is indeed a classical H~{\sc ii} region. 
The str\"{o}mgren radius was calculated to be 0.051 $\pm$ 0.002 pc 
and the dynamical timescale to be 1.07 $\pm$ 0.31 Myr.  
\item
The grism slitless spectroscopic image was used to identify the H$\alpha$ emission line stars;
NIR CC diagram (\textit{J-H/H-K}) to identify the Class II-type objects, and the NIR CM diagram 
(\textit{K/H-K}) to identify the 
red sources ($H-K > 0.6$) in our field. The average extinction in the field was calculated to be 2.70 mag. 
\item
We estimated the age and mass of the YSOs using our optical CM diagram (\textit{V/V-I}), besides using the  
NIR mass spectrum to estimate the mass range. 
Age was found to be in the range 0.5 to 2 Myr, with the average age being $\sim$ 1 Myr.
The mass was estimated to be $\sim$ 0.1 to 2 M$_{\odot}$, with a few outliers $\sim$ 3 M$_{\odot}$. 
\item
We studied the evolutionary status of some of the sources from \citet{lin10},
red sources, Class II-type objects, and H$\alpha$ emission line stars using the SED model of \citet{rob07}. 
The models predict - for most of the sources - stellar age in the range 0.05$-$1 Myr, stellar 
mass in the range 0.2$-$2.5 M$_{\odot}$, and disk accretion rate $\sim$ 10$^{-7}-$10$^{-11}$ M$_{\odot}$yr$^{-1}$.
A few H$\alpha$ emission line stars, which are most evolved YSOs in the field, have stellar ages upto 5 Myr, 
and masses $\sim$ 2.5$-$5 M$_{\odot}$. 
\item
The spatial distribution of the YSOs indicates a possible evolutionary sequence, in the 
sense that youngest sources are distributed away from the ionizing source. 
This could be evidence in  
support of triggered star formation in the region.  
The star formation seems to have propagated from the ionizing source towards the 
cloud \object{LDN1657A}, which is further supported by the presence of massive protostar UYSO1 
discovered by \citet{for04}, and cool, compact sources discovered by \citet{lin10}, which are located 
away from the ionizing source.
\end{enumerate}

We thank the anonymous referee for a very
thorough and critical reading of our manuscript. The useful comments and
suggestions made by the referee greatly improved the scientific content of
the paper.
The authors thank the staff of HCT, operated by Indian Institute
of Astrophysics (Bangalore); IGO at Girawali, operated by 
Inter University Centre for Astronomy \& Astrophysics (IUCAA), Pune; 
IRSF at South Africa in joint partnership between S.A.A.O and 
Nagoya University of Japan; and GMRT managed by National Center 
for Radio Astrophysics of the Tata Institute of Fundamental 
Research (TIFR) for their assistance and support during observations.
This paper used data from the NRAO VLA Archive Survey (NVAS). 
The NVAS can be accessed through http://www.aoc.nrao.edu/$\sim$vlbacald/.
MT is supported by KAKENHI 22000005. KKM acknowledges support from a Marie Curie IRSES
grant (230843) under the auspices of which some part of this work was carried out.

{}

\clearpage
\LongTables

\begin{deluxetable}{c c c c c c c c c c c}
\tablecolumns{11}
\tablewidth{0pt}
\tablenum{2}
\tabletypesize{\tiny}
\renewcommand{\arraystretch}{0.8}
\tablecaption{Optical \& infrared magnitudes for the YSOs \label{table_YSOs}}
\tablehead{
\colhead{ID}  &   \colhead{RA}   &    \colhead{Dec.}      &      \colhead{\textit{V}}     &   \colhead{\textit{I}}   &   \colhead{\textit{J}}    &       
 \colhead{\textit{H}}    &   \colhead{\textit{K$_{\rm S}$}}  &    \colhead{w1}    &      \colhead{w2}     &      \colhead{w3}   \\
\colhead{} & \colhead{(J2000)} & \colhead{(J2000)} & \colhead{} & \colhead{} & \colhead{}  & \colhead{} & \colhead{}
 & \colhead{} & \colhead{} & \colhead{}}
\startdata
\multicolumn{11}{c}{LDN1657A-2 \& 3 from Linz et al. (2010)} \\
\cline{1-11} 
 1  &   07:04:58.030  &   -12:16:50.80  &              \nodata  &              \nodata  &              \nodata  &              \nodata  &              \nodata  &    14.12 $\pm$  0.08  &    10.78 $\pm$  0.02  &     8.62 $\pm$  0.09  \\
 2  &   07:05:00.665  &   -12:16:45.13  &              \nodata  &              \nodata  &    16.85 $\pm$  0.10  &    14.67 $\pm$  0.05  &    12.96 $\pm$  0.04  &    11.12 $\pm$  0.01  &     9.73 $\pm$  0.01  &     7.17 $\pm$  0.02  \\
\cline{1-11}
\multicolumn{11}{c}{YSOs with $H-K>0.6$} \\
\cline{1-11}
 3  &   07:04:55.818  &   -12:17:37.90  &              \nodata  &              \nodata  &              \nodata  &    16.62 $\pm$  0.03  &    15.51 $\pm$  0.09  &              \nodata  &              \nodata  &              \nodata  \\
 4  &   07:04:56.305  &   -12:17:24.14  &              \nodata  &              \nodata  &              \nodata  &    16.82 $\pm$  0.03  &    15.47 $\pm$  0.07  &    14.78 $\pm$  0.11  &    13.69 $\pm$  0.10  &              \nodata  \\
 5  &   07:04:56.334  &   -12:16:14.28  &              \nodata  &              \nodata  &              \nodata  &    17.32 $\pm$  0.04  &    16.18 $\pm$  0.13  &              \nodata  &              \nodata  &              \nodata  \\
 6  &   07:04:56.580  &   -12:15:14.47  &              \nodata  &              \nodata  &              \nodata  &    16.67 $\pm$  0.02  &    15.80 $\pm$  0.10  &    14.63 $\pm$  0.08  &              \nodata  &              \nodata  \\
 7  &   07:04:56.634  &   -12:16:59.69  &              \nodata  &              \nodata  &    17.39 $\pm$  0.04  &    14.95 $\pm$  0.01  &    13.60 $\pm$  0.02  &    12.64 $\pm$  0.03  &              \nodata  &              \nodata  \\
 8  &   07:04:56.898  &   -12:16:39.00  &              \nodata  &              \nodata  &              \nodata  &    15.87 $\pm$  0.07  &    15.09 $\pm$  0.06  &    13.71 $\pm$  0.06  &    12.49 $\pm$  0.08  &              \nodata  \\
 9  &   07:04:57.702  &   -12:16:20.06  &              \nodata  &              \nodata  &              \nodata  &    13.32 $\pm$  0.01  &    12.21 $\pm$  0.01  &    11.06 $\pm$  0.02  &    10.91 $\pm$  0.02  &              \nodata  \\
10  &   07:04:57.927  &   -12:20:36.37  &              \nodata  &              \nodata  &              \nodata  &    17.14 $\pm$  0.05  &    16.12 $\pm$  0.13  &              \nodata  &              \nodata  &              \nodata  \\
11  &   07:04:58.407  &   -12:17:47.23  &              \nodata  &              \nodata  &              \nodata  &    16.60 $\pm$  0.03  &    15.27 $\pm$  0.06  &    14.78 $\pm$  0.10  &    14.21 $\pm$  0.16  &              \nodata  \\
12  &   07:04:58.749  &   -12:16:18.35  &              \nodata  &              \nodata  &              \nodata  &    16.27 $\pm$  0.03  &    15.15 $\pm$  0.06  &              \nodata  &              \nodata  &              \nodata  \\
13  &   07:04:59.150  &   -12:15:17.78  &              \nodata  &              \nodata  &              \nodata  &    16.73 $\pm$  0.03  &    16.05 $\pm$  0.14  &              \nodata  &              \nodata  &              \nodata  \\
14  &   07:04:59.994  &   -12:15:41.34  &              \nodata  &              \nodata  &    17.65 $\pm$  0.03  &    16.12 $\pm$  0.02  &    15.34 $\pm$  0.06  &              \nodata  &              \nodata  &              \nodata  \\
15  &   07:05:00.319  &   -12:16:43.75  &              \nodata  &              \nodata  &              \nodata  &    15.69 $\pm$  0.03  &    13.44 $\pm$  0.02  &              \nodata  &              \nodata  &              \nodata  \\
16  &   07:05:00.599  &   -12:18:52.50  &              \nodata  &              \nodata  &    17.56 $\pm$  0.03  &    16.25 $\pm$  0.02  &    15.52 $\pm$  0.08  &              \nodata  &              \nodata  &              \nodata  \\
17  &   07:05:00.826  &   -12:18:46.47  &              \nodata  &              \nodata  &              \nodata  &    16.80 $\pm$  0.05  &    16.12 $\pm$  0.13  &              \nodata  &              \nodata  &              \nodata  \\
18  &   07:05:00.881  &   -12:18:26.03  &              \nodata  &              \nodata  &    17.55 $\pm$  0.05  &    16.86 $\pm$  0.04  &    16.10 $\pm$  0.14  &    14.59 $\pm$  0.10  &              \nodata  &              \nodata  \\
19  &   07:05:00.919  &   -12:15:42.37  &    19.65 $\pm$  0.01  &    15.29 $\pm$  0.01  &    12.47 $\pm$  0.01  &    10.83 $\pm$  0.01  &    10.07 $\pm$  0.01  &     9.61 $\pm$  0.01  &     9.47 $\pm$  0.01  &              \nodata  \\
20  &   07:05:01.313  &   -12:17:22.48  &              \nodata  &              \nodata  &              \nodata  &    16.83 $\pm$  0.03  &    15.45 $\pm$  0.08  &    14.72 $\pm$  0.10  &              \nodata  &              \nodata  \\
21  &   07:05:01.873  &   -12:17:11.83  &              \nodata  &              \nodata  &              \nodata  &    16.95 $\pm$  0.04  &    15.87 $\pm$  0.12  &    14.82 $\pm$  0.16  &              \nodata  &              \nodata  \\
22  &   07:05:02.311  &   -12:17:17.60  &              \nodata  &              \nodata  &              \nodata  &    16.99 $\pm$  0.03  &    16.21 $\pm$  0.13  &              \nodata  &              \nodata  &              \nodata  \\
23  &   07:05:02.637  &   -12:16:44.26  &              \nodata  &              \nodata  &              \nodata  &    16.29 $\pm$  0.02  &    14.99 $\pm$  0.05  &              \nodata  &    14.33 $\pm$  0.24  &              \nodata  \\
24  &   07:05:02.833  &   -12:16:12.72  &              \nodata  &              \nodata  &              \nodata  &    17.34 $\pm$  0.04  &    15.72 $\pm$  0.09  &    14.46 $\pm$  0.18  &    13.69 $\pm$  0.10  &              \nodata  \\
25  &   07:05:03.666  &   -12:18:24.20  &              \nodata  &              \nodata  &              \nodata  &    16.09 $\pm$  0.02  &    14.27 $\pm$  0.03  &    13.23 $\pm$  0.05  &    12.15 $\pm$  0.05  &              \nodata  \\
26  &   07:05:03.832  &   -12:18:00.40  &              \nodata  &              \nodata  &    16.64 $\pm$  0.02  &    15.10 $\pm$  0.01  &    14.35 $\pm$  0.03  &    13.12 $\pm$  0.06  &              \nodata  &              \nodata  \\
27  &   07:05:04.448  &   -12:17:09.32  &              \nodata  &              \nodata  &              \nodata  &    15.98 $\pm$  0.01  &    14.31 $\pm$  0.03  &    13.26 $\pm$  0.05  &    12.57 $\pm$  0.05  &              \nodata  \\
28  &   07:05:05.984  &   -12:17:21.69  &              \nodata  &              \nodata  &              \nodata  &    16.88 $\pm$  0.03  &    15.37 $\pm$  0.06  &              \nodata  &              \nodata  &              \nodata  \\
29  &   07:05:05.995  &   -12:16:45.27  &              \nodata  &              \nodata  &              \nodata  &    17.15 $\pm$  0.04  &    15.39 $\pm$  0.07  &    14.65 $\pm$  0.14  &              \nodata  &              \nodata  \\
30  &   07:05:06.044  &   -12:17:17.39  &              \nodata  &              \nodata  &              \nodata  &    16.91 $\pm$  0.04  &    15.65 $\pm$  0.09  &              \nodata  &              \nodata  &              \nodata  \\
31  &   07:05:06.075  &   -12:18:35.45  &              \nodata  &              \nodata  &              \nodata  &    16.55 $\pm$  0.03  &    14.59 $\pm$  0.03  &    13.27 $\pm$  0.09  &    12.06 $\pm$  0.07  &              \nodata  \\
32  &   07:05:06.216  &   -12:17:33.02  &              \nodata  &              \nodata  &    18.25 $\pm$  0.07  &    16.46 $\pm$  0.03  &    15.48 $\pm$  0.08  &              \nodata  &              \nodata  &              \nodata  \\
33  &   07:05:06.735  &   -12:17:52.68  &              \nodata  &              \nodata  &              \nodata  &    16.85 $\pm$  0.03  &    15.85 $\pm$  0.11  &              \nodata  &              \nodata  &              \nodata  \\
34  &   07:05:07.139  &   -12:19:23.87  &              \nodata  &              \nodata  &    17.59 $\pm$  0.05  &    15.48 $\pm$  0.01  &    14.34 $\pm$  0.03  &              \nodata  &              \nodata  &              \nodata  \\
35  &   07:05:08.827  &   -12:17:54.66  &              \nodata  &              \nodata  &    17.69 $\pm$  0.03  &    16.31 $\pm$  0.02  &    15.60 $\pm$  0.08  &              \nodata  &              \nodata  &              \nodata  \\
36  &   07:05:08.890  &   -12:15:34.44  &              \nodata  &              \nodata  &              \nodata  &    16.89 $\pm$  0.06  &    16.00 $\pm$  0.13  &    15.72 $\pm$  0.27  &              \nodata  &              \nodata  \\
37  &   07:05:09.181  &   -12:19:25.46  &              \nodata  &              \nodata  &    16.00 $\pm$  0.01  &    14.15 $\pm$  0.01  &    13.26 $\pm$  0.01  &              \nodata  &              \nodata  &              \nodata  \\
38  &   07:05:09.283  &   -12:17:24.01  &              \nodata  &              \nodata  &    17.70 $\pm$  0.04  &    16.31 $\pm$  0.02  &    15.63 $\pm$  0.09  &              \nodata  &              \nodata  &              \nodata  \\
39  &   07:05:09.326  &   -12:18:42.24  &              \nodata  &              \nodata  &    14.38 $\pm$  0.04  &    12.84 $\pm$  0.06  &    12.05 $\pm$  0.05  &    10.88 $\pm$  0.06  &    10.12 $\pm$  0.06  &              \nodata  \\
40  &   07:05:09.749  &   -12:20:02.05  &              \nodata  &              \nodata  &    17.45 $\pm$  0.04  &    16.22 $\pm$  0.02  &    15.52 $\pm$  0.08  &              \nodata  &              \nodata  &              \nodata  \\
41  &   07:05:09.811  &   -12:19:26.43  &              \nodata  &              \nodata  &              \nodata  &    16.97 $\pm$  0.04  &    16.06 $\pm$  0.14  &              \nodata  &              \nodata  &              \nodata  \\
42  &   07:05:10.071  &   -12:18:51.91  &              \nodata  &              \nodata  &              \nodata  &    18.02 $\pm$  0.10  &    16.19 $\pm$  0.14  &              \nodata  &              \nodata  &              \nodata  \\
43  &   07:05:10.124  &   -12:18:29.40  &              \nodata  &              \nodata  &    16.65 $\pm$  0.01  &    14.53 $\pm$  0.01  &    13.40 $\pm$  0.02  &              \nodata  &              \nodata  &              \nodata  \\
44  &   07:05:10.358  &   -12:18:09.88  &              \nodata  &              \nodata  &    17.46 $\pm$  0.03  &    16.26 $\pm$  0.02  &    15.61 $\pm$  0.09  &              \nodata  &              \nodata  &              \nodata  \\
45  &   07:05:10.395  &   -12:19:25.70  &              \nodata  &              \nodata  &              \nodata  &    16.64 $\pm$  0.06  &    15.66 $\pm$  0.10  &    11.20 $\pm$  0.10  &    10.85 $\pm$  0.12  &     5.12 $\pm$  0.07  \\
46  &   07:05:10.481  &   -12:18:30.70  &              \nodata  &              \nodata  &              \nodata  &    17.59 $\pm$  0.07  &    15.89 $\pm$  0.10  &              \nodata  &              \nodata  &              \nodata  \\
47  &   07:05:11.475  &   -12:19:16.58  &              \nodata  &              \nodata  &              \nodata  &    16.55 $\pm$  0.06  &    15.38 $\pm$  0.13  &              \nodata  &              \nodata  &              \nodata  \\
48  &   07:05:11.946  &   -12:18:48.14  &              \nodata  &              \nodata  &    15.46 $\pm$  0.01  &    13.99 $\pm$  0.01  &    13.32 $\pm$  0.01  &              \nodata  &              \nodata  &              \nodata  \\
49  &   07:05:13.678  &   -12:19:51.96  &              \nodata  &              \nodata  &    15.89 $\pm$  0.01  &    14.52 $\pm$  0.01  &    13.83 $\pm$  0.02  &    11.81 $\pm$  0.13  &    11.10 $\pm$  0.10  &              \nodata  \\
50  &   07:05:17.596  &   -12:20:23.48  &              \nodata  &              \nodata  &              \nodata  &    16.88 $\pm$  0.04  &    16.14 $\pm$  0.13  &              \nodata  &              \nodata  &              \nodata  \\
51  &   07:05:17.668  &   -12:20:26.39  &              \nodata  &              \nodata  &    17.47 $\pm$  0.04  &    16.66 $\pm$  0.03  &    15.86 $\pm$  0.11  &              \nodata  &              \nodata  &              \nodata  \\
52  &   07:05:23.035  &   -12:19:56.15  &              \nodata  &              \nodata  &    17.03 $\pm$  0.02  &    16.28 $\pm$  0.02  &    15.64 $\pm$  0.08  &              \nodata  &              \nodata  &              \nodata  \\
\cline{1-11}
\multicolumn{11}{c}{Class II-type sources} \\
\cline{1-11}
53  &   07:04:56.673  &   -12:16:12.77  &    20.19 $\pm$  0.03  &    16.94 $\pm$  0.01  &    15.81 $\pm$  0.01  &    14.25 $\pm$  0.01  &    13.29 $\pm$  0.02  &    12.04 $\pm$  0.02  &    11.43 $\pm$  0.02  &              \nodata  \\
54  &   07:04:57.964  &   -12:17:13.85  &              \nodata  &              \nodata  &    16.08 $\pm$  0.01  &    15.07 $\pm$  0.01  &    14.37 $\pm$  0.04  &    14.14 $\pm$  0.06  &    12.92 $\pm$  0.06  &              \nodata  \\
55  &   07:04:59.938  &   -12:17:24.19  &              \nodata  &              \nodata  &    17.61 $\pm$  0.05  &    15.61 $\pm$  0.03  &    14.43 $\pm$  0.04  &    13.17 $\pm$  0.04  &    12.94 $\pm$  0.07  &              \nodata  \\
56  &   07:05:00.844  &   -12:20:06.48  &              \nodata  &              \nodata  &    17.20 $\pm$  0.04  &    16.32 $\pm$  0.02  &    15.70 $\pm$  0.09  &              \nodata  &              \nodata  &              \nodata  \\
57  &   07:05:01.117  &   -12:16:50.42  &              \nodata  &              \nodata  &    17.35 $\pm$  0.04  &    16.08 $\pm$  0.03  &    15.13 $\pm$  0.06  &              \nodata  &              \nodata  &              \nodata  \\
58  &   07:05:01.230  &   -12:19:03.69  &    20.20 $\pm$  0.03  &    16.73 $\pm$  0.01  &    14.83 $\pm$  0.01  &    13.60 $\pm$  0.01  &    12.85 $\pm$  0.01  &    11.77 $\pm$  0.02  &    11.11 $\pm$  0.02  &              \nodata  \\
59  &   07:05:02.673  &   -12:15:44.75  &              \nodata  &              \nodata  &    16.70 $\pm$  0.02  &    15.84 $\pm$  0.01  &    15.28 $\pm$  0.07  &              \nodata  &              \nodata  &              \nodata  \\
60  &   07:05:03.364  &   -12:15:53.58  &              \nodata  &              \nodata  &    17.78 $\pm$  0.04  &    16.75 $\pm$  0.03  &    16.09 $\pm$  0.14  &              \nodata  &              \nodata  &              \nodata  \\
61  &   07:05:04.206  &   -12:18:45.13  &              \nodata  &              \nodata  &    17.90 $\pm$  0.06  &    15.30 $\pm$  0.01  &    13.80 $\pm$  0.02  &    12.44 $\pm$  0.04  &    11.37 $\pm$  0.03  &              \nodata  \\
62  &   07:05:04.644  &   -12:19:10.39  &              \nodata  &              \nodata  &    15.96 $\pm$  0.02  &    14.82 $\pm$  0.01  &    14.13 $\pm$  0.02  &    13.46 $\pm$  0.13  &              \nodata  &              \nodata  \\
63  &   07:05:05.166  &   -12:18:19.49  &              \nodata  &              \nodata  &    17.43 $\pm$  0.04  &    15.88 $\pm$  0.02  &    14.82 $\pm$  0.04  &              \nodata  &              \nodata  &              \nodata  \\
64  &   07:05:08.269  &   -12:19:15.83  &              \nodata  &              \nodata  &    16.16 $\pm$  0.01  &    13.69 $\pm$  0.01  &    12.16 $\pm$  0.01  &    10.41 $\pm$  0.06  &     9.03 $\pm$  0.03  &              \nodata  \\
65  &   07:05:08.771  &   -12:18:51.26  &              \nodata  &              \nodata  &    16.67 $\pm$  0.02  &    15.41 $\pm$  0.02  &    14.64 $\pm$  0.04  &              \nodata  &              \nodata  &              \nodata  \\
66  &   07:05:09.197  &   -12:19:00.00  &              \nodata  &              \nodata  &    17.09 $\pm$  0.03  &    15.29 $\pm$  0.01  &    13.95 $\pm$  0.02  &    11.16 $\pm$  0.08  &    10.86 $\pm$  0.17  &              \nodata  \\
67  &   07:05:10.638  &   -12:19:15.02  &              \nodata  &              \nodata  &    17.04 $\pm$  0.02  &    15.89 $\pm$  0.03  &    15.10 $\pm$  0.06  &              \nodata  &              \nodata  &              \nodata  \\
68  &   07:05:10.818  &   -12:19:43.90  &              \nodata  &              \nodata  &    17.16 $\pm$  0.04  &    15.83 $\pm$  0.02  &    14.89 $\pm$  0.05  &              \nodata  &              \nodata  &              \nodata  \\
69  &   07:05:11.424  &   -12:19:24.36  &              \nodata  &              \nodata  &    16.93 $\pm$  0.04  &    15.13 $\pm$  0.01  &    13.76 $\pm$  0.04  &    11.04 $\pm$  0.10  &    10.45 $\pm$  0.10  &              \nodata  \\
70  &   07:05:11.541  &   -12:18:11.39  &    20.02 $\pm$  0.05  &    16.58 $\pm$  0.03  &    14.63 $\pm$  0.01  &    13.09 $\pm$  0.01  &    12.15 $\pm$  0.01  &              \nodata  &    10.04 $\pm$  0.08  &              \nodata  \\
71  &   07:05:11.642  &   -12:19:30.95  &              \nodata  &              \nodata  &    15.10 $\pm$  0.01  &    13.96 $\pm$  0.01  &    13.14 $\pm$  0.01  &              \nodata  &              \nodata  &              \nodata  \\
72  &   07:05:11.774  &   -12:16:24.01  &    20.78 $\pm$  0.05  &    17.95 $\pm$  0.02  &    15.34 $\pm$  0.01  &    13.72 $\pm$  0.01  &    12.52 $\pm$  0.01  &    11.27 $\pm$  0.02  &    10.29 $\pm$  0.02  &     7.89 $\pm$  0.07  \\
73  &   07:05:14.495  &   -12:20:47.57  &              \nodata  &              \nodata  &    15.25 $\pm$  0.01  &    14.16 $\pm$  0.01  &    13.47 $\pm$  0.02  &    12.43 $\pm$  0.12  &    11.65 $\pm$  0.12  &              \nodata  \\
74  &   07:05:14.610  &   -12:19:42.29  &              \nodata  &              \nodata  &    15.07 $\pm$  0.01  &    13.62 $\pm$  0.01  &    12.74 $\pm$  0.01  &              \nodata  &              \nodata  &              \nodata  \\
75  &   07:05:15.324  &   -12:20:02.05  &              \nodata  &              \nodata  &    17.17 $\pm$  0.04  &    15.89 $\pm$  0.01  &    14.95 $\pm$  0.05  &    11.61 $\pm$  0.22  &    11.30 $\pm$  0.27  &              \nodata  \\
76  &   07:05:17.419  &   -12:18:58.69  &              \nodata  &              \nodata  &    14.42 $\pm$  0.01  &    13.49 $\pm$  0.01  &    12.88 $\pm$  0.01  &    11.95 $\pm$  0.04  &    11.21 $\pm$  0.06  &              \nodata  \\
77  &   07:05:18.113  &   -12:19:25.47  &              \nodata  &              \nodata  &    17.25 $\pm$  0.03  &    15.90 $\pm$  0.01  &    14.94 $\pm$  0.05  &              \nodata  &              \nodata  &              \nodata  \\
78  &   07:05:18.950  &   -12:20:13.53  &              \nodata  &              \nodata  &    16.03 $\pm$  0.05  &    14.93 $\pm$  0.03  &    13.97 $\pm$  0.03  &              \nodata  &     9.47 $\pm$  0.05  &              \nodata  \\
\cline{1-11}
\multicolumn{11}{c}{H$\alpha$ emission line sources} \\
\cline{1-11}
79  &   07:05:06.330  &   -12:15:36.38  &    17.43 $\pm$  0.01  &    14.82 $\pm$  0.01  &    12.95 $\pm$  0.01  &    12.01 $\pm$  0.01  &    11.33 $\pm$  0.01  &    10.44 $\pm$  0.01  &     9.84 $\pm$  0.01  &     7.36 $\pm$  0.03  \\
80  &   07:05:09.652  &   -12:19:56.32  &    20.07 $\pm$  0.03  &    16.18 $\pm$  0.01  &    14.05 $\pm$  0.01  &    12.77 $\pm$  0.01  &    12.02 $\pm$  0.01  &    10.91 $\pm$  0.07  &    10.21 $\pm$  0.07  &              \nodata  \\
81  &   07:05:10.920  &   -12:15:22.46  &              \nodata  &              \nodata  &    15.11 $\pm$  0.01  &    14.49 $\pm$  0.01  &    14.11 $\pm$  0.03  &    13.61 $\pm$  0.04  &    13.12 $\pm$  0.05  &              \nodata  \\
82  &   07:05:12.292  &   -12:18:38.27  &    16.74 $\pm$  0.01  &    14.37 $\pm$  0.01  &    13.01 $\pm$  0.01  &    12.06 $\pm$  0.01  &    11.50 $\pm$  0.01  &    11.05 $\pm$  0.11  &    10.53 $\pm$  0.12  &              \nodata  \\
83  &   07:05:15.359  &   -12:19:39.02  &              \nodata  &              \nodata  &    13.13 $\pm$  0.06  &    12.27 $\pm$  0.06  &    11.99 $\pm$  0.05  &    10.93 $\pm$  0.12  &    10.27 $\pm$  0.10  &              \nodata  \\
84  &   07:05:15.840  &   -12:19:45.09  &    15.41 $\pm$  0.14  &    14.39 $\pm$  0.02  &    13.41 $\pm$  0.01  &    12.55 $\pm$  0.01  &    12.10 $\pm$  0.01  &              \nodata  &              \nodata  &              \nodata  \\
85  &   07:05:16.265  &   -12:20:17.61  &              \nodata  &              \nodata  &    14.32 $\pm$  0.01  &    13.37 $\pm$  0.01  &    12.84 $\pm$  0.01  &              \nodata  &              \nodata  &              \nodata  \\
86  &   07:05:17.904  &   -12:15:40.26  &    15.27 $\pm$  0.01  &    14.17 $\pm$  0.01  &    13.89 $\pm$  0.01  &    13.65 $\pm$  0.02  &    13.40 $\pm$  0.03  &    12.86 $\pm$  0.03  &    12.50 $\pm$  0.04  &              \nodata  \\
87  &   07:05:18.494  &   -12:15:09.22  &    17.35 $\pm$  0.01  &    14.85 $\pm$  0.01  &    14.00 $\pm$  0.01  &    12.85 $\pm$  0.01  &    11.95 $\pm$  0.01  &    10.77 $\pm$  0.01  &     9.97 $\pm$  0.01  &     8.21 $\pm$  0.03  \\
88  &   07:05:19.389  &   -12:20:28.71  &    14.37 $\pm$  0.01  &    12.92 $\pm$  0.01  &    11.83 $\pm$  0.01  &    11.11 $\pm$  0.01  &    10.79 $\pm$  0.01  &    10.09 $\pm$  0.06  &     9.50 $\pm$  0.05  &              \nodata  \\
89  &   07:05:19.592  &   -12:17:31.48  &    16.82 $\pm$  0.01  &    14.59 $\pm$  0.01  &    13.38 $\pm$  0.01  &    12.51 $\pm$  0.01  &    12.10 $\pm$  0.01  &    11.40 $\pm$  0.01  &    10.74 $\pm$  0.02  &     7.82 $\pm$  0.03  \\
90  &   07:05:20.068  &   -12:19:12.59  &    13.86 $\pm$  0.01  &    12.69 $\pm$  0.01  &    11.67 $\pm$  0.01  &    10.92 $\pm$  0.01  &    10.40 $\pm$  0.01  &     9.82 $\pm$  0.01  &     9.36 $\pm$  0.01  &     6.62 $\pm$  0.04  \\
91  &   07:05:20.728  &   -12:21:42.69  &    17.11 $\pm$  0.01  &    15.14 $\pm$  0.01  &    13.83 $\pm$  0.01  &    12.84 $\pm$  0.01  &    12.15 $\pm$  0.01  &    11.03 $\pm$  0.02  &    10.38 $\pm$  0.02  &              \nodata  \\
92  &   07:05:21.912  &   -12:19:07.19  &    19.80 $\pm$  0.02  &    15.92 $\pm$  0.01  &    14.30 $\pm$  0.01  &    13.44 $\pm$  0.01  &    13.01 $\pm$  0.01  &    12.53 $\pm$  0.03  &    12.02 $\pm$  0.03  &              \nodata  \\
93  &   07:05:22.359  &   -12:17:18.63  &    18.23 $\pm$  0.01  &    15.39 $\pm$  0.01  &    14.17 $\pm$  0.01  &    13.30 $\pm$  0.01  &    12.84 $\pm$  0.01  &    12.25 $\pm$  0.02  &    11.62 $\pm$  0.03  &              \nodata  \\
\cline{1-11}
\multicolumn{11}{c}{H$\alpha$ emission line sources outside NIR FoV\tablenotemark{a}} \\
\cline{1-11}
94  &   07:05:12.139  &   -12:14:41.57  &              \nodata  &              \nodata  &    14.15 $\pm$  0.04  &    13.26 $\pm$  0.04  &    12.77 $\pm$  0.03  &    12.26 $\pm$  0.02  &    11.76 $\pm$  0.02  &     9.38 $\pm$  0.03  \\
95  &   07:05:16.154  &   -12:14:55.90  &    18.57 $\pm$  0.01  &    15.73 $\pm$  0.01  &    14.24 $\pm$  0.03  &    13.30 $\pm$  0.03  &    12.77 $\pm$  0.03  &    12.15 $\pm$  0.02  &    11.53 $\pm$  0.02  &    10.00 $\pm$  0.09  \\  
96  &   07:05:23.669  &   -12:22:36.08  &    16.18 $\pm$  0.01  &    14.14 $\pm$  0.01  &    13.25 $\pm$  0.03  &    12.48 $\pm$  0.02  &    12.16 $\pm$  0.02  &    11.89 $\pm$  0.02  &    11.46 $\pm$  0.03  &     8.33 $\pm$  0.10  \\  
97  &   07:05:37.457  &   -12:21:17.32  &              \nodata  &              \nodata  &    13.98 $\pm$  0.02  &    13.14 $\pm$  0.02  &    12.73 $\pm$  0.03  &    12.04 $\pm$  0.05  &    11.34 $\pm$  0.07  &              \nodata  \\  
\enddata
\tablenotetext{a}{NIR magnitudes taken from 2MASS catalogue}
\end{deluxetable}

\clearpage

\end{document}